\documentclass[lettersize,journal]{IEEEtran}
\usepackage{amsmath,amsfonts}
\usepackage{array}
\usepackage[caption=false,font=scriptsize,labelfont=rm,textfont=rm]{subfig}
\usepackage{textcomp}
\usepackage{stfloats}
\usepackage{threeparttable}
\usepackage{url}
\usepackage{verbatim}
\usepackage{graphicx}
\def\BibTeX{{\rm B\kern-.05em{\sc i\kern-.025em b}\kern-.08em
		T\kern-.1667em\lower.7ex\hbox{E}\kern-.125emX}}
\usepackage{balance}

\usepackage{multirow}
\usepackage{cite}
\usepackage{cases}
\usepackage{amssymb}
\usepackage{algorithm}
\usepackage{algpseudocode}
\usepackage{bm}
\usepackage{color,xcolor}

\usepackage{float}    
\usepackage{subfig}   
\usepackage{overpic}  

\usepackage{amsthm}

\newtheorem{mypro}{Proposition}

\allowdisplaybreaks[3]

\begin{document}
	
	\title{Covert Communications in MEC-Based Networked ISAC Systems Towards Low-Altitude Economy}
	\author{Weihao Mao,~\IEEEmembership{Student Member,~IEEE}, Yang Lu,~\IEEEmembership{Member,~IEEE}, 
    Bo Ai,~\IEEEmembership{Fellow,~IEEE}, 
    \\ and Tony Q. S. Quek,~\IEEEmembership{Fellow,~IEEE}
		\thanks{Weihao Mao and Yang Lu are with the State Key Laboratory of Advanced Rail Autonomous Operation, and also with the School of Computer Science and Technology, Beijing Jiaotong University, Beijing 100044, China (e-mail: weihaomao@bjtu.edu.cn, yanglu@bjtu.edu.cn).}
        \thanks{Bo Ai is with the School of Electronics and Information Engineering, Beijing Jiaotong University, Beijing 100044, China (e-mail: boai@bjtu.edu.cn).}
        \thanks{Tony Q. S. Quek is with the Information Systems Technology and Design Pillar, Singapore University of Technology and Design, Singapore 487372 (e-mail: tonyquek@sutd.edu.sg).}
        \thanks{This work has been submitted to the IEEE for possible publication. Copyright may be transferred without notice, after which this version may no longer be accessible.}
	}
    
	\maketitle
	\thispagestyle{empty}
	
	\begin{abstract}
        Low-altitude economy (LAE) is an emerging business model, which heavily relies on integrated sensing and communications (ISAC), mobile edge computing (MEC), and covert communications. This paper investigates the convert transmission design in MEC-based networked ISAC systems towards LAE, where an MEC server coordinates multiple access points to simultaneously receive computation tasks from multiple unmanned aerial vehicles (UAVs), locate a target in a sensing area, and maintain UAVs' covert transmission against multiple wardens. We first derive closed-form expressions for the detection error probability (DEP) at wardens. Then, we formulate a total energy consumption minimization problem by optimizing  communication, sensing, and computation resources as well as  UAV trajectories, subject to the requirements on quality of MEC services, DEP, and radar signal-to-interference-and-noise ratio, and  the causality of UAV trajectories. An alternating optimization based algorithm is proposed to handle the considered problem, which decomposes it into two subproblems: joint optimization of communication, sensing, and computation resources, and  UAV trajectory optimization. The former is addressed by a successive convex approximation based algorithm, while the latter is solved via a trust-region based algorithm. Simulations validate the effectiveness of the proposed algorithm compared with various benchmarks, and reveal the trade-offs among communication, sensing, and computation in LAE systems. 
	\end{abstract}
	
	
	\IEEEpeerreviewmaketitle

	\setlength{\parindent}{1em}
	
	\section{Introduction}

    Low-altitude economy (LAE) \cite{bc_LAE_def} is an emerging business model driven by various manned and unmanned aircraft, such as electric vertical take-off and landing aircraft \cite{bc_evtl} and unmanned aerial vehicles (UAVs) \cite{bc_uav}, engaged in low-altitude (typically below $1,000~{\rm m}$ \cite{bc_low_al}) flight activities. Owing to the advantages of low-altitude aircraft in terms of operational flexibility, dynamic mobility, and real-time data acquisition capabilities \cite{bc_LAE_adv}, LAE has effectively enabled diverse applications, including environmental monitoring and emergency medical rescue \cite{bc_LAE_application}. Wireless communication systems play a crucial role in enabling high-quality information exchange among aircraft. Nevertheless,  as one of the most promising future paradigms, LAE imposes additional requirements on sensing, computation, and information privacy for existing wireless communication systems. First, LAE requires  ubiquitous sensing to enable safe trajectory planning for numerous aircraft while maintaining real-time monitoring to prevent collision with unauthorized targets \cite{bc_LAE_ISAC}. Second, LAE demands that aircraft possess either onboard computation capability or reliable access to proximal computing nodes to efficiently handle emergent computation-intensive and latency-sensitive tasks \cite{bc_LAE_MEC}. Third, LAE demands enhanced covertness of information transmission due to the exposed nature of low-altitude wireless channels in the presence of  wardens \cite{bc_LAE_covert}. To meet these requirements, the development of LAE increasingly relies on novel B5G techniques, i.e., integrated sensing and communications (ISAC), mobile edge computing (MEC), and covert communications. Thus far, there have been some existing works focusing on the integration of these techniques with LAE,  detailed as follows. 

    As a key candidate technology for B5G wireless systems, ISAC integrates traditionally separated base station (BS) communication and radar sensing into the same time-frequency resource block through joint waveform design \cite{bc_ISAC_def}. Due to its ability to simultaneously reduce hardware costs \cite{bc_ISAC_adv1}, improve spectrum efficiency \cite{bc_ISAC_adv2}, and  generate synergistic gains for both communication and sensing \cite{bc_ISAC_adv3}, ISAC has been widely adopted in LAE systems, see  e.g., \cite{rw_ISAC1,rw_ISAC2,rw_ISAC3,rw_ISAC4, rw_ISAC5, rw_ISAC6}. In \cite{rw_ISAC1} and \cite{rw_ISAC2}, the authors investigated the UAV-enabled ISAC systems with UAV acting an aerial BS, and maximized the weighted sum rate of users, subject to the sensing requirements,  where the beam pattern gain and Cramér-Rao bound (CRB) were utilized as metrics, respectively. Moreover, \cite{rw_ISAC3} investigated a hybrid-reconfigurable intelligent surface (RIS) empowered UAV-assisted ISAC system with UAV being an aerial RIS, and maximized the sum rate under the constraints of sensing rate. In addition to mono-static ISAC studied in \cite{rw_ISAC1, rw_ISAC2, rw_ISAC3}, networked ISAC leveraging the coordination among multiple BSs are gaining increasing attention in LAE systems, see e.g., \cite{rw_ISAC4, rw_ISAC5, rw_ISAC6}. In \cite{rw_ISAC4}, the authors proposed a two-stage framework to estimate the  parameters of UAVs via the cooperation among multiple ground BSs (GBSs). Besides, \cite{rw_ISAC5} introduced a cooperative networked ISAC framework for the UAV mobility management in LAE systems, where a primary GBS and two secondary GBSs jointly performed tracking and handover operations during inter-cell transitions. Furthermore, \cite{rw_ISAC6} maximized the weighted sum rate under the constraints of minimum illumination power to the sensing area in networked ISAC systems, where multiple GBSs were coordinated to provide information service to UAVs. 

    Meanwhile, MEC serves as a critical enabler for processing unpredictable computation-intensive and latency-sensitive tasks from intelligent mobile applications \cite{bc_MEC_def}. Compared to  traditional cloud computing architectures \cite{bc_MEC_cloud}, MEC deploys computational resources at network edge nodes in closer proximity to users, achieving significant latency reduction for users \cite{bc_MEC_adv}. This paradigm has motivated some research on MEC in LAE systems, see e.g., \cite{rw_MEC1, rw_MEC2, rw_MEC3, rw_MEC4, rw_MEC5, rw_MEC6}. The authors in \cite{rw_MEC1} and \cite{rw_MEC2} considered the aerial MEC system empowered by single UAV, aiming to achieve the maximum energy efficiency of UAV and the min-max computation latency among users, respectively. Moreover, \cite{rw_MEC3} coordinated multiple UAVs equipped with MEC servers to minimize the energy consumption required to complete computation tasks within the specified latency. However, the authors in \cite{rw_MEC1, rw_MEC2, rw_MEC3} exclusively employed UAV-mounted MEC servers, which  may face inherent limitations in processing large-scale computation tasks due to constrained computing capability and finite battery capacity at UAVs. To overcome these challenges, recent works have proposed hybrid UAV-MEC architectures, which integrate terrestrial MEC servers into UAV-assisted MEC systems, see e.g., \cite{rw_MEC4, rw_MEC5, rw_MEC6}. In \cite{rw_MEC4}, the authors employed the UAV as an aerial relay to forward computation tasks from users to the GBS equipped with powerful MEC server with the aim to maximize the secure computing capacity. Furthermore, \cite{rw_MEC5} and \cite{rw_MEC6} minimized the energy consumption of the UAV and the weighted summation of energy consumption and task delay in single-UAV and multi-UAV  MEC systems, respectively, where UAVs function as both MEC servers and relays.

    In parallel, covert communications become the central consideration of information privacy for  future wireless networks, attracting significant attention from both academic and industry \cite{bc_covert_def1}. This technology aims to conceal communication activities from detection by watchful wardens \cite{bc_covert_def2}. However, the implementation of covert communications in LAE systems faces significant challenges. This is because the open nature and high quality line-of-sight (LoS) characteristics of air-to-air and air-to-ground channels may be exploited by wardens for transmission detection \cite{bc_covert_chall}. Therefore, some existing works focus on covert communications in LAE systems, see  e.g., \cite{rw_covert1, rw_covert2, rw_covert3, rw_covert4, rw_covert5}. In \cite{rw_covert1}, the communication signal-to-interference-and-noise ratio (SINR) was maximized in a UAV-assisted covert communication system, where multiple UAVs were utilized as jammers. Furthermore, \cite{rw_covert2} maximized the communication rate in UAV-RIS assisted covert communications systems where the UAV served as an aerial RIS. Most recently, several studies have integrated ISAC or MEC capabilities into UAV covert communications, see  e.g., \cite{rw_covert3, rw_covert4, rw_covert5}.  In \cite{rw_covert3}, the authors minimized the maximum task latency in an MEC-based covert communications system, where two UAVs served as an aerial MEC server and a jammer, respectively. Besides, \cite{rw_covert4} and \cite{rw_covert5} maximized the minimum and sum communication rate in UAV-assisted integrated sensing and covert communications systems with radar signal-to-noise ratio (SNR) and beam pattern gain as metrics, respectively. 
    

    \emph{To the authors' best knowledge, the joint resource allocation design of covert communications, sensing, and computation for LAE systems has not been reported in the open literature yet.}  It is worth noting that ISAC, MEC, and covert communications constitute three essential enabling technologies for LAE systems, which can address collision avoidance, support large-scale computing, and protect information privacy, respectively. However, developing an efficient resource allocation scheme for such integrated systems faces two fundamental challenges: 1) the tight coupling among communication, sensing, and computation resources, and 2) the inherent difficulty of maintaining covertness in low-altitude environments with strong LoS conditions. \emph{To fill this gap}, we formulate an MEC-based networked ISAC system towards LAE, and propose an alternating optimization (AO)-based algorithm to support covert communications. The main contributions of this paper are summarized as follows.
     \begin{enumerate}
        \item 
        We formulate a covertness-aware MEC-based networked ISAC system towards LAE, where an MEC server coordinates multiple access points (APs) to simultaneously receive computation tasks from multiple UAVs, locate one potential aerial target, and ensure covert transmission of UAVs against multiple aerial wardens. Notably, the waveforms generated by APs is dual-functional, i.e., jamming signals to interfere with wardens and sensing signals to locate the target.

        \item
        We first derive closed-form expressions for the detection error probability (DEP) at wardens under optimal decision threshold conditions, and then transform the DEP into a tractable form. Subsequently, we formulate a total energy consumption minimization problem subject to quality of service (QoS) requirements of MEC services, DEP requirements for covert communications, radar SINR requirements, and the causality of UAV trajectories, by jointly optimizing the communication, sensing, and computation resources as well as  UAV trajectories. 

        \item
        To address the considered problem, an AO-based algorithm is proposed to decompose it into two subproblems, i.e., joint optimization of communication, sensing, and computation resources, and UAV trajectory optimization. For the former, we first introduce a series of auxiliary variables to reformulate it, and propose  a successive convex approximation (SCA)-based algorithm to handle the reformulated subproblem. For the latter, we propose a trust-region-based algorithm to solve it iteratively via first-order Taylor expansion.

        \item 
        Simulations are provided to evaluate the proposed algorithm and reveal the trade-offs among communication, sensing, and computation resources in LAE systems. The results illustrate that the UAVs tends to evade wardens to prevent both physical collisions and information leakage. Besides, the proposed algorithm achieves the best performance among several benchmarks including straight flight design, power allocation design, fixed time assignment design, and full-offloading design.        
    \end{enumerate}

    {\it Notations:} In this paper, the symbols $x$, ${\bf x}$, ${\bf X}$, and ${\cal X}$ represent a scalar, a vector, a matrix, and a set, respectively. ${\rm Re}(\cdot)$ denote the real part of a complex number, vector, or matrix. $|| \cdot ||$ denotes the Euclidean norm for a complex vector. ${\rm Tr}( \cdot )$ denotes the trace for a complex matrix. $(\cdot)^T$ and $(\cdot)^H$ denote the transpose, and conjugate transpose, respectively. ${\mathbb C}^M$ and  ${\mathbb C}^{M \times N}$ denote the set of $M \times 1$ complex-valued vectors and $M \times N$ complex-valued matrices, respectively. ${\bf X} \succeq {\bf 0}$ denotes ${\bf X}$ is a positive semi-definite matrix. ${\bf a} \sim \mathcal{CN}({\bm \mu}, {\bm \Sigma})$ denotes that ${\bf a}$ is a circularly symmetric complex Gaussian random variable with mean  ${\bm \mu}$ and covariance matrix ${\bm \Sigma}$. $\mathbb{E}\{\cdot \}$ and ${\rm Prob}\{ \cdot \}$ denote the expectation function and probability function, respectively.

	\section{System Model}
	\begin{figure}
		\begin{center}
			\centerline{\includegraphics[ width=.49\textwidth]{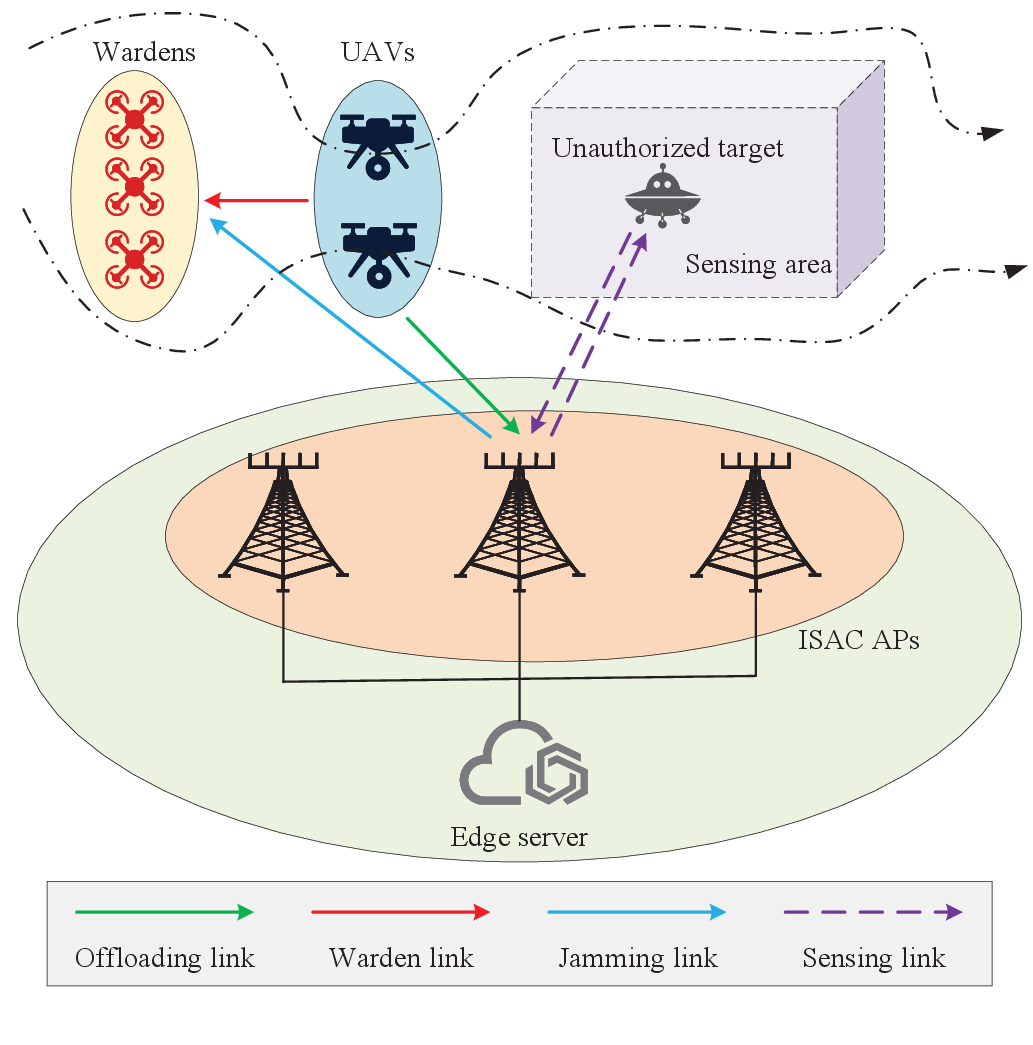}}
			\caption{Illustration of an MEC-based networked ISAC system.}
			\label{sys}
		\end{center}
	\end{figure}

	We consider an MEC-based networked ISAC system as shown in Fig. \ref{sys}. In the system, $M$ ISAC APs are responsible for receiving  computation tasks offloaded by $K$ UAVs in the presence of $L$ passive aerial wardens, and forwarding the tasks to an MEC server with high-performance computing capabilities.  Meanwhile, the ISAC APs intend to locate an unauthorized target within a sensing area. Notably, the waveform generated by the ISAC APs is dual-functional. First, it serves as jamming signals to prevent wardens from detecting whether UAVs are offloading tasks. Second, it covers the sensing area to perform target localization. We assume that each of $M$ ISAC APs is equipped with $N_{\rm T}$ transmit uniform linear antennas (ULAs) and $N_{\rm R}$ receive ULAs,  $K$ UAVs are $N_{\rm U}$-ULA, and $L$ wardens are single-antenna.   For clarity, let $\mathcal{M} \triangleq \{1,2,\ldots, M\}$, $\mathcal{K} \triangleq \{1,2,\ldots, K\}$, and $\mathcal{L} \triangleq \{1,2,\ldots, L\}$ denote the index sets of ISAC APs, UAVs, and wardens, respectively.

	
	The MEC-based ISAC service lasts for a duration $T$, which is divided into $N$ time slots based on the discrete path planning approach \cite{discrete_path}, and each time slot lasts $\Delta_{\rm T} \triangleq T/N$. Here, $N$ is assumed to be large enough such that $\Delta_{\rm T}$ is sufficiently small and the UAV locations can be assumed to be unchanged during each time slot. Without loss of generality, a three-dimension Cartesian coordinate system is built, where coordinates of the $m$-th AP and $l$-th warden are denoted by ${\bf q}_m^{\rm AP} = [x_m^{\rm AP}, y_m^{\rm AP}, 0]^T$ and  ${\bf q}_l^{\rm W} = [x_l^{\rm W}, y_l^{\rm W}, H_l^{\rm W}]^T$, respectively. In the $n$-th time slot ($n \in \mathcal{N} \triangleq \{1,2, \ldots, N\}$), coordinates of the $k$-th UAV and the target are denoted by ${\bf q}_k^{\rm U}[n] = [{\bf u}_k^T[n], H_k]^T$ and ${\bf q}_{\rm T}[n] = [x_{\rm T}[n], y_{\rm T}[n], H_{\rm T}[n] ]^T$, respectively, where ${\bf u}_k[n] = [x_k^{\rm U}[n], y_k^{\rm U}[n]]^T$ denotes the horizontal coordinate of the $k$-th UAV to be optimized and $H_k$ denotes the pre-defined flight altitude of the $k$-th UAV. During the $n$-th time slot, the propulsion power consumption of the $k$-th UAV is given by
	\begin{flalign}
		P(||{\bf v}_k [n]||)  &~=  P_0 \left(  1 + \frac{3 ||{\bf v}_k[n] ||^2}{U_{\rm tip}^2}    \right) + \frac{1}{2} d_0 \rho_0 s A || {\bf v}_k[n] ||^3 \nonumber	\\
		& + P_{\rm H} \left(\sqrt{1 + \frac{ ||{\bf v}_k[n] ||^4 }{4 v_0^4}} - \frac{||{\bf v}_k[n]||^2}{2 v_0^2}  \right)^{\frac{1}{2} },
	\end{flalign}
	where $P_0$ and $P_{\rm H}$ respectively denote the blade profile power and induced power, $U_{\rm tip}$ denotes the tip speed of the rotor induced velocity, $d_0$ denotes the fuselage drag ratio, $\rho_0$ denotes the air density, $s$ is the rotor solidity, $A$ is the rotor disc area, $v_0$ is the mean rotor induced velocity, and ${\bf v}_k[n]$ is the horizontal velocity of the $k$-th UAV, which is calculated by
	\begin{flalign*}
		{\bf v}_k [n]  =  \frac{{\bf u}_k[n+1] - {\bf u}_k[n]  } {\Delta_{\rm T}}.
	\end{flalign*}
	
	\subsection{Channel Model}
	The considered system comprises four types of wireless links: 1) offloading links from UAVs to ISAC APs, 2) warden links from UAVs to wardens, 3) jamming links from ISAC APs to wardens, and  4) cascaded sensing links upon which ISAC APs transmit signals to the target and receive the reflected echoes. Similar with existing works, such as \cite{los1, los2, los3}, these links are assumed to be LoS as they represent either air-ground  or air-air transmission. The channel models of the four links are detailed as follows.
    
\subsubsection{Offloading links} The channel from the $k$-th UAV to the $m$-th ISAC AP at the $n$-th time slot is given by
	\begin{flalign*}
		&{\bf H}_{m, k} [n]  = \nonumber \\
		  &\sqrt{ \frac{C_0}{ || {\bf q}_k^{\rm U}[n]  - {\bf q}_m^{\rm AP} ||^2   }} {\bf a}_{\rm R} \left( \theta_{m, k}[n]  \right) {\bf a}_{\rm U}^T\left( \theta_{m, k}[n]   \right)\in{\mathbb C}^{N_{\rm R}\times N_{\rm U}},
	\end{flalign*}
	where $C_0$, ${\bf a}_{\rm R}(\theta_{m, k}[n])$, and ${\bf a}_{\rm U} ({\theta}_{m, k}[n])$ denote the path-loss at the reference distance, the receive steering vector at the $m$-th AP, and the transmit steering vector at the $k$-th UAV, respectively. These steering vectors are respectively denoted by
	\begin{flalign*}
		& {\bf a}_{\rm R} \left( \theta_{m, k}[n] \right) = \left[1, \ldots,   e^{j 2 \pi \frac{d}{\lambda} \cos(\theta_{m,k}[n])(N_{\rm R}-1) }    \right]^T\in{\mathbb C}^{N_{\rm R}},                               \\
		& {\bf a}_{\rm U}\left( \theta_{m, k}[n]  \right) =  \left[1, \ldots, e^{j 2\pi \frac{d}{\lambda} \cos(\theta_{m, k}[n]) (N_{\rm U}  - 1) }      \right]^T\in{\mathbb C}^{N_{\rm U}}, 
	\end{flalign*} 
	where $\theta_{m, k}[n]$ denotes the angle of departure (AoD) from the $k$-th UAV to the $m$-th AP, i.e., 
	\begin{flalign*}
		\theta_{m, k}[n] = \arccos \frac{H_k}{\left\| {\bf q}_k^{\rm U}[n] - {\bf q}_m^{\rm AP}    \right\|}.
	\end{flalign*}
	
\subsubsection{Warden links} The channel from the $k$-th UAV to the $l$-th warden at the $n$-th time slot is given by
	\begin{flalign*}
		{\bf h}_{l, k}[n] = \sqrt{\frac{C_0}{ \left\| {\bf q}_k^{\rm U}[n] - {\bf q}_l^{\rm W}  \right\|^2  }} {\bf a}_{\rm U} \left( \phi_{l,k}[n]  \right)\in{\mathbb C}^{N_{\rm U}},
	\end{flalign*}
	where $\phi_{l, k}[n]$ denotes the AoD from the $k$-th UAV to the $l$-th warden, i.e.,
	\begin{flalign*}
		\phi_{l,k}[n] = \arccos \frac{H_l^{\rm W} - H_k}{ \left\|{\bf q}_k^{\rm U}[n] - {\bf q}_l^{\rm W}   \right\|  }.
	\end{flalign*}

\subsubsection{Jamming links} The channel from the $m$-th ISAC AP to the $l$-th warden at the $n$-th time slot is given by
	\begin{flalign*}
		{\bf g}_{m, l} = \sqrt{\frac{C_0}{\left\|{\bf q}_m^{\rm AP} - {\bf q}_l^{\rm W}  \right\|^2}} {\bf a}_{\rm T} \left( \alpha_{m, l}  \right)\in{\mathbb C}^{ N_{\rm T}},
	\end{flalign*}
	where ${\bf a}_{\rm T}( \alpha_{m, l})$ denotes the transmit steering vector at the $m$-th AP, which is calculated by 
	\begin{flalign*}
		{\bf a}_{\rm T} \left(\alpha_{m, l} \right) = \left[ 1, \ldots, e^{j 2 \pi \frac{d}{\lambda} \cos( \alpha_{m, l}) (N_{\rm T} - 1)  } \right]^T\in{\mathbb C}^{N_{\rm T}},
	\end{flalign*}
	with $\alpha_{m, l}$ denoting the AoD from the $m$-th AP to the $l$-th warden, i.e.,
	\begin{flalign*}
		\alpha_{m, l} = \arccos \frac{H_l^{\rm W}}{ || {\bf q}_m^{\rm AP} - {\bf q}_l^{\rm W} ||   }.
	\end{flalign*}

\subsubsection{Cascaded sensing links} The $\langle m, j \rangle$-th AP-target-AP cascaded channel is given by
	\begin{flalign*}
		{\bf G}_{m, j} [n] = \frac{C_0 {\bf a}_{\rm R} \left( \beta_j[n] \right) {\bf a}_{\rm T}^T \left( \beta_m[n] \right)   }{||{\bf q}_m^{\rm AP} - {\bf q}_{\rm T}[n]  || \times || {\bf q}_j^{\rm AP} - {\bf q}_{\rm T}[n]  ||  }\in{\mathbb C}^{N_{\rm R}\times N_{\rm T}},
	\end{flalign*}
	where $\beta_m[n]$ is the AoD from the $m$-th AP to the target, i.e., 
	\begin{flalign*}
		\beta_m [n] = \arccos\frac{H_{\rm T}[n]}{||{\bf q}_m^{\rm AP} - {\bf q}_{\rm T}[n]  ||}.
	\end{flalign*}
	
	\subsection{Signal Processing Model}
	
	\begin{figure}
		\begin{center}
			\centerline{\includegraphics[ width=.49\textwidth]{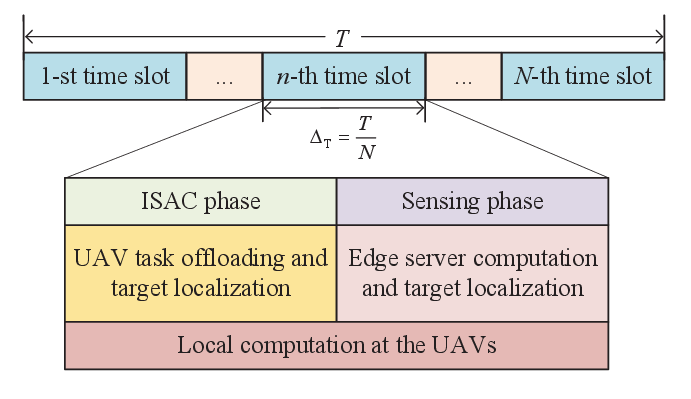}}
			\caption{Illustration of time slot division.}
			\label{time_slot}
		\end{center}
	\end{figure}
	
     Fig. \ref{time_slot} illustrates the time slot assignment at UAVs and APs. Throughout each time slot, UAVs conduct local computation. However, ISAC APs divide their time slots into two phases: 1) offloading-sensing phase, where APs receive the computation tasks offloaded by UAVs while performing target localization, and 2) {computing-sensing phase}, where the MEC server executes the offloaded computation tasks while APs continue target localization.

\begin{figure*}
    \begin{flalign}
		&{\bf y}_{\rm R}[n] = \begin{cases}
		  {\bf G}[n] {\bf W}_{\rm s}^0[n] {\bf s}_{\rm s}[n] + {\bf n}_{\rm R}[n],&{\rm Phase~2},  \\
		  \sum_{i \in \mathcal{K}} {\bf H}_i[n] {\bf w}_i[n] s_i[n] + {\bf G}[n] {\bf W}_{\rm s}^{1}[n]  {\bf s}_{\rm s}[n] + {\bf n}_{\rm R}[n], ~~&{\rm Phase~1}.
		\end{cases} \label{y_R_n}
	\end{flalign}
    \hrule
    \end{figure*}
    
\subsubsection{Signal propagation model} For UAVs, we denote  $s_i[n] \sim \mathcal{CN}(0, 1)$ and ${\bf w}_i[n] \in \mathbb{C}^{N_{\rm u}}$ by the desired information symbol transmitted by the $i$-th UAV and the corresponding beamforming vector. For APs, we denote 
    ${\bf s}_{\rm s}[n] \sim \mathcal{CN}({\bf 0}, {\bf I}_{M N_{\rm T}} )$, ${\bf W}_{\rm s}^0[n] \in \mathbb{C}^{MN_{\rm T} \times MN_{\rm T}},$ and ${\bf W}_{\rm s}^1[n] \in  \mathbb{C}^{MN_{\rm T} \times MN_{\rm T}}$ by the dedicated sensing symbol, the corresponding sensing beamforming matrices in the two phases, respectively. Then, the aggregated signal received at the MEC server from all APs is defined by ${\bf y}_{\rm R}[n]$, which is expressed as \eqref{y_R_n}, where  ${\bf G}[n] $ denotes the aggregated channel of AP-target-AP sensing links, i.e.,
	 \begin{flalign*}
	 	{\bf G}[n] \triangleq 
	 	\begin{bmatrix}
	 		{\bf G}_{1,1}[n]  & \cdots  & {\bf G}_{M, 1}[n] \\
	 		\vdots  & \ddots & \vdots \\
	 		{\bf G}_{1, M}[n] & \cdots & {\bf G}_{M, M}[n]
	 	\end{bmatrix}\in \mathbb{C}^{M N_{\rm R} \times M N_{\rm T}},
	 \end{flalign*}
	and ${\bf H}_i[n]  $ denotes the aggregated channel from the $i$-th UAV to the MEC server with
	\begin{flalign*}
		{\bf H}_i[n] \triangleq \left[{\bf H}_{1, i}[n]; {\bf H}_{2, i}[n]; \ldots; {\bf H}_{M, i}[n]    \right]\in \mathbb{C}^{MN_{\rm R}\times N_{\rm U}}.
	\end{flalign*}
	 ${\bf n}_{\rm R}[n] \sim \mathcal{CN}({\bf 0}, \sigma_{\rm R}^2{\bf I}_{M N_{\rm R}} )$ denotes the additive white Gaussian noise (AWGN) at the MEC server with $\sigma_{\rm R}^2$ being the noise power.


	\subsubsection{Communication model}

   With ${\bf y}_{\rm R}[n]$ in Phase 1, the $k$-th UAV offloads the bits associated with computation tasks  to the MEC server via APs, with number of
	\begin{flalign}\label{l_o_k}
		l_{{\rm o}, k}[n] = t_1[n] B \log_2 \left(1 + \Gamma_k[n]  \right),
	\end{flalign}
	where $t_1[n]\le\Delta_{\rm T}$ denotes the time allocated to the offloading-sensing phase, $B$ denotes the transmission bandwidth, and $\Gamma_k[n]$ denotes SINR of the $k$-th UAV, which is given by \eqref{sinrk} with ${\bf R}_{\rm s}^1[n] \triangleq {\bf W}_{\rm s}^1[n] ({\bf W}_{\rm s}^1)^H[n]  $.
	\begin{figure*}
		\begin{flalign}
			\Gamma_k[n] = \frac{ {\rm Tr}\left( {\bf H}_k[n] {\bf w}_k[n] {\bf w}_k^H[n] {\bf H}_k^H[n]  \right)  }{{\rm Tr} \left( \sum_{i \in \mathcal{K} \setminus \{k\} } {\bf H}_i[n] {\bf w}_i[n] {\bf w}_i^H[n] {\bf H}_i^H[n] + {\bf G}[n] {\bf R}_{\rm s}^1[n] {\bf G}^H[n]    \right) + M N_{\rm R} \sigma_{\rm R}^2 }      
			\label{sinrk}
		\end{flalign}
		\hrule
	\end{figure*}

    The energy consumption during the offloading-sensing phase for $K$ UAVs to offload tasks is given by
    \begin{flalign}
        E_{\rm C}[n] = t_1[n] \sum\nolimits_{k \in \mathcal{K}} {\bf w}_k^H[n] {\bf w}_k[n].
    \end{flalign}

	\subsubsection{Sensing model}
The APs conduct target sensing in the two phases with different sensing beamformers. The MEC server utilize the echo signals (i.e., ${\bf y}_{\rm R}[n]$ in Phase 1 and Phase 2) for target location. The radar SINR at the MEC server is expressed as
\begin{flalign}
		&\Gamma_{\rm r} [n] = \\
        &		\begin{cases}
			\frac{{\rm Tr}\left({\bf G}[n] {\bf R}_{\rm s}^0[n] {\bf G}^H[n]  \right) }{ M N_{\rm R} \sigma_{\rm R}^2},&{\rm Phase~2},   \\
			\frac{{\rm Tr}\left( {\bf G}[n] {\bf R}_{\rm s}^1[n] {\bf G}^H[n]   \right)  }{\sum_{i \in \mathcal{K}}  {\rm Tr}\left( {\bf H}_i[n] {\bf w}_i[n] {\bf w}_i^H[n] {\bf H}_i^H[n] \right) + M N_{\rm R} \sigma_{\rm R}^2 },&{\rm Phase~1},  
		\end{cases} \nonumber
	\end{flalign}
    where ${\bf R}_{\rm s}^0[n] \triangleq {\bf W}_{\rm s}^0[n] ({\bf W}_{\rm s}^0[n])^H$.

    For practical implementation, the MEC server selects $Q$ location samples within the sensing area to form a set of $\mathcal{Q}$. Then, to guarantee sensing performance, the expectation of radar SINR, defined by the weighted mean of received radar SINR over the time slot, is required to exceed a pre-defined threshold $\Gamma_{\min}$ for any sampled point within $\mathcal{Q}$, i.e.,
	\begin{flalign}
		&\mathbb{E}\left\{ \Gamma_{\rm r} \right[n]\} = \frac{t_0[n]}{\Delta_{\rm T}} \frac{ {\rm Tr} \left( {\bf G}[n] {\bf R}_{\rm s}^0[n] {\bf G}^H[n]  \right)   }{M N_{\rm R} \sigma_{\rm R}^2  } \nonumber \\
		&+  \frac{t_1[n]}{\Delta_{\rm T}} \frac{{\rm Tr}\left( {\bf G}[n] {\bf R}_{\rm s}^1[n] {\bf G}^H[n]   \right)  }{\sum_{i \in \mathcal{K}} {\rm Tr}\left( {\bf H}_i[n] {\bf w}_i[n] {\bf w}_i^H[n] {\bf H}_i^H[n] \right)  + M N_{\rm R} \sigma_{\rm R}^2    } \nonumber \\
		&\geq \Gamma_{\min},~\forall~{\bf q}_{\rm T}[n] \in \mathcal{Q}, \label{con_sen}
	\end{flalign}
	where {$t_0[n]\le(\Delta_{\rm T}-t_1[n])$} denotes the time allocated to the computing-sensing phase.

    The energy consumption for $M$ APs to sense the target is given by
    \begin{flalign}
        E_{\rm S}[n] = t_0[n] {\rm Tr}\left( {\bf R}_{\rm s}^0[n]  \right) + t_1[n] {\rm Tr} \left( {\bf R}_{\rm s}^1[n]  \right).
    \end{flalign}

	\subsubsection{Computation model} Denote $I_k[n]$ as the bits of the computation tasks of the $k$-th UAV. The computation task of the $k$-th UAV is partitioned into two components, i.e., local computation at the UAV and edge computation at the MEC server. 
    
    For local computing, denote $f_{{\rm l}, k}[n]$ as the computation resource (in CPU cycles/s) of the $k$-th UAV to handle its computation tasks. Then, the bits of computation task finished at the $k$-th UAV is given by
	\begin{flalign}
		l_{{\rm l}, k}[n] = \frac{f_{{\rm l}, k}[n]}{D_k} \Delta_{\rm T},
	\end{flalign}
	where $D_k$ denotes the required number of CPU cycles to process one bit of the computation task. Consequently, the energy consumption due to local computation is given by \cite{mec_power}
	\begin{flalign}
		E_{{\rm l}, k}[n] = v_{\rm l} f_{{\rm l}, k}^3[n] \Delta_{\rm T},
	\end{flalign}
	where $v_{\rm l}$ denotes the effective capacitance coefficient of the processor. The task partition scheme requires the offloaded bits (cf. \eqref{l_o_k}) to satisfy
	\begin{flalign}
		 l_{{\rm o}, k}[n] \geq I_k[n] - l_{{\rm l}, k}[n],~ \forall{\bf q}_{\rm T}[n] \in \mathcal{Q}. \label{con_com1}
	\end{flalign}
   Note that $\forall{\bf q}_{\rm T}[n] \in \mathcal{Q}$ arises from the interaction between communication and sensing during the offloading-sensing phase.
	
	For edge computation, denote $f_{{\rm u}, k}[n]$ as  the computation resource of the MEC server allocated to the $k$-th UAV. Then, the bits of the computation task of the $k$-th UAV finished at the MEC server is given by
	\begin{flalign}
		l_{{\rm u}, k}[n] = \frac{ f_{{\rm u}, k}[n] }{D_k} t_0[n],
	\end{flalign}
	which results in corresponding energy consumption as
	\begin{flalign}
		E_{{\rm u}, k}[n] = v_{\rm u} f_{{\rm u}, k}^3[n] t_0[n], 
	\end{flalign}
	where $v_{\rm u}$ denotes the effective capacitance coefficient of the processor at the MEC server. To guarantee the computation task to be finished, the following condition must be satisfied: 
	\begin{flalign}
		 l_{{\rm u}, k}[n]  \geq I_k[n] - l_{{\rm l}, k}[n] .\label{con_com2}
	\end{flalign}
    
    After the computation tasks are processed at the MEC server, the size of corresponding results tends to be small. Consequently, the time duration for the MEC server to feedback results to UAVs and the associated energy consumption are assumed to be negligible \cite{without_res_time}. 
	
	\subsection{Covert Communication Model}
As depicted in Fig. \ref{time_slot}, the wardens perform transmission detection for each time slot. 
At the $n$-th time slot, we consider two hypotheses, i.e., ${\cal H}_0$ represents the null hypothesis corresponding to the computing-sensing phase, and ${\cal H}_1$ represents the alternative hypothesis representing the offloading-sensing phase. Under the two hypotheses, the received signals at the $l$-th warden are given by
	\begin{flalign}
		&y_{{\rm W}, l}[n] = \label{ywl} \\
		& \nonumber
		\begin{cases}
		{\bf g}_l^H  {\bf W}_{\rm s}^0[n] {\bf s}_{\rm s}[n]  + n_l[n],   &   {\cal H}_0,  \\
		\sum_{i \in \mathcal{K}} {\bf h}_{l, i}^H[n] {\bf w}_i[n] s_i[n] + {\bf g}_l^H {\bf W}_{\rm s}^1[n] {\bf s}_{\rm s}[n] +  n_l[n],   	  &   {\cal H}_1, 
		\end{cases}
	\end{flalign}
	where ${\bf g}_l$ denotes the aggregated channel from all APs to the $l$-th warden with
	\begin{flalign*}
		{\bf g}_l = \left[{\bf g}_{1, l}; {\bf g}_{2, l}; \ldots;  {\bf g}_{M, l}  \right] \in \mathbb{C}^{M N_{\rm T}} ,
	\end{flalign*} 
	and $n_l[n] \sim \mathcal{CN}(0, \sigma_l^2)$ denotes the AWGN at the $l$-th warden with $\sigma_l^2$ being the noise power. 
	
	Based on the Neyman-Pearson criterion and Neyman-Fisher factorization theorem \cite{Neyman-Pearson}, the $l$-th warden determines whether UAVs are offloading computation tasks based on the following likelihood ratio test:
	\begin{flalign}
		|y_{{\rm W}, l}[n]|^2  \underset{{\rm D}_1}{\overset{{\rm D}_0}{\lessgtr}} \delta_l[n],
	\end{flalign}
	where $\delta[n]$ denotes the pre-defined decision threshold, and ${\rm D}_0$ and ${\rm D}_1$ denote the decisions for hypotheses ${\cal H}_0$ and ${\cal H}_1$, respectively. According to \eqref{ywl}, the expectations of the received signal power at the $l$-th warden under the two hypotheses are
	\begin{flalign}
		\mathbb{E} \left\{ \left| y_{{\rm W}, l}[n]   \right|^2   \right\} = 
		\begin{cases}
		\lambda_l^0[n],	         & {\cal H}_0, \\
		\lambda_l^1[n],	         & {\cal H}_1,
		\end{cases}
	\end{flalign}
	where 
	\begin{flalign*}
		\begin{cases}
			\lambda_l^0[n] = {\bf g}_l^H {\bf R}_{\rm s}^0[n] {\bf g}_l + \sigma_l^2 ,  \\
			\lambda_l^1[n]  = \sum\limits_{i \in \mathcal{K}} {\bf h}_{l, i}^H[n] {\bf w}_i[n] {\bf w}_i^H[n] {\bf h}_{l, i}[n] + {\bf g}_l^H {\bf R}_{\rm s}^1[n] {\bf g}_l + \sigma_l^2.
		\end{cases}
	\end{flalign*}
	
	This paper considers the worst-case scenario where the $l$-th warden employs the optimal decision threshold to minimize the DEP $\xi_l[n]$, which is given by
	\begin{flalign}
		 \xi_l[n] = {\rm Prob}\left\{{\rm D}_1 | {\cal H}_0, \delta_l[n]  \right\} + {\rm Prob}\left\{{\rm D}_0 | {\cal H}_1, \delta_l[n]   \right\}.
	\end{flalign}
	The optimal $\delta_l^{\star}[n]$ and the corresponding $\xi_l^{\star}[n]$ are given in the following proposition.
	
	\begin{mypro}
		The worst-case scenario indicates that the optimal decision threshold at the $l$-th warden in the $n$-th time slot is given by
		\begin{flalign}
			\delta_l^{\star}[n] = \lambda_l^{0}[n] \frac{1 + \mu_l[n] }{\mu_l[n]} \ln\left( 1+\mu_l[n] \right),
		\end{flalign}
		where $\mu_l[n]$ is given by
		\begin{flalign}
			& \mu_l[n] =  \\
			& \frac{\sum\nolimits_{i \in \mathcal{K}} {\bf h}_{l, i}^H[n] {\bf w}_i[n] {\bf w}_i^H[n] {\bf h}_{l, i}[n] + {\bf g}_l^H \left(  {\bf R}_{\rm s}^1[n] -  {\bf R}_{\rm s}^0[n]  \right) {\bf g}_l }{ {\bf g}_l^H {\bf R}_{\rm s}^0[n] {\bf g}_l + \sigma_l^2  } . \nonumber
		\end{flalign}
		Subsequently, the corresponding minimum DEP is given by
		\begin{flalign}
			\xi_l^{\star}[n] =  1 + e^{-\frac{1 + \mu_l[n] }{\mu_l[n]} \ln(1 +\mu_l[n]) } - e^{ -\frac{1}{\mu_l[n]} \ln(1+\mu_l[n]) }. \label{xi_star}
		\end{flalign}
	\end{mypro}
	\begin{IEEEproof}
		The proof follows Appendix A in \cite{rw_covert5} with modifications only to the expressions for $\lambda_l^0[n]$ and $\lambda_l^1[n]$. Due to the space limitation, the detailed proof is omitted. Please refer to \cite{rw_covert5}.
	\end{IEEEproof}
	
	Regarding covert transmission, DEP at each warden must be larger than a threshold, i.e.,
	\begin{flalign}
	 \xi_l^{\star}[n] \geq \xi_{\min} \overset{(a)}{\iff}
		 1 -\xi_{\min}   \geq F( \mu_l[n] ),   \label{dep1}
	\end{flalign}
	where (a) is due to \eqref{xi_star}, and $F(\mu_l[n]) \triangleq \mu_l[n](1 + \mu_l[n])^{-(1+1/\mu_l[n])} $. It is observed that 
	\begin{flalign}
		\frac{{\rm d} \ln F\left( \mu_l[n]  \right)  }{{\rm d} \mu_l[n]} = \frac{ \ln(1 + \mu_l[n ]) }{ \mu_l^2[n]  } > 0,
	\end{flalign}
	which indicates that $F(\mu_l[n])$ is monotonically increasing with respect to (w.r.t.) $\mu_l[n]$. Thus, the DEP constraint \eqref{dep1} can be rewritten as
	\begin{flalign}
		\mu_l[n] \leq F^{-1} (1 - \xi_{\min}) \triangleq \mu_{\max}. \label{con_covert}
	\end{flalign}
	
	\subsection{Problem Formulation}
	We aim to minimize the total energy consumption during the service period $T$ subject to the QoS requirements of MEC service, radar SINR requirements, covert transmission requirements, the causality of trajectories of UAVs, and available transmission and computation resource  by jointly optimizing the transmit beamformers at both UAVs and ISAC APs, the computation resource allocation, the time division scheme, and the trajectories of UAVs. The considered optimization problem is formulated as:
	\begin{subequations}
		\begin{flalign}
			{{\bf P}_0}: & \mathop{\min}  \limits_{ {\bf L}_0 }   \sum\nolimits_{n \in \mathcal{N}}\Big( E_{\rm C}[n] + E_{\rm S}[n]  + \sum\nolimits_{k \in \mathcal{K}} \left( P(||{\bf v}_k [n]||)  \Delta_{\rm T} \right.  \nonumber \\
            &  \left.
            ~~~~~~~~~~~~~~~+ E_{{\rm l}, k}[n] + E_{{\rm u}, k}[n]    \right) \Big)
			  \label{p0a}  \\
			{\rm s.t.} ~
		&  0 \leq f_{{\rm l}, k}[n] \leq f_{{\rm l}, \max}, \label{p0b} \\ 
        & \sum\nolimits_{k \in \mathcal{K}} f_{{\rm u}, k} [n]\leq f_{{\rm u}, \max} \label{p0c_v1} \\
		& {\bf w}_k^H[n] {\bf w}_k[n] \leq P_{{\rm U}, \max},
		\label{p0c}  \\
		& {\rm Tr}\left({\bf R}_{\rm s}^0[n]  \right) \leq P_{{\rm AP}, \max},~ {\rm Tr}\left({\bf R}_{\rm s}^1[n]  \right) \leq P_{{\rm AP}, \max}, 
		\label{p0d} \\
		& t_0[n] + t_1[n] \leq \Delta_{\rm T}, \label{p0e} \\
		& f_{{\rm u}, k}[n] \geq 0,~t_0[n] \geq 0,~t_1[n] \geq 0,
        \label{p0f} \\
		& {\bf R}_{\rm s}^0[n] \succeq {\bf 0},~{\bf R}_{\rm s}^1[n] \succeq {\bf 0}, 
        \label{p0g} \\
		& || {\bf u}_k[n] - {\bf u}_k[n-1] || \leq V_{\max} \Delta_{\rm T} \label{p0i} \\
		& {\bf u}_k[0] = {\bf u}_k^{\rm I}, {\bf u}_k[N] = {\bf u}_k^{\rm F}, \label{p0j} \\
		& ||{\bf u}_k[n] - {\bf u}_i[n]||^2 + (H_k-H_i)^2 \geq D_{\min}^2, \label{p0k} \\
		& || {\bf q}_k^{\rm U}[n] - {\bf q}_l^{\rm W}  ||^2 \geq D_{\min}^2, \label{p0l} \\
		&|| {\bf q}_k^{\rm U}[n] - {\bf q}_{\rm T}[n]  ||^2 \geq D_{\min}^2, \label{p0m} \\
		& \eqref{con_sen}, \eqref{con_com1}, \eqref{con_com2}, \eqref{con_covert}, \nonumber \\
		& \forall {\bf q}_{\rm T}[n] \in  \mathcal{Q},~\forall i, k \in \mathcal{K}, i \neq k,~\forall l \in \mathcal{L},~\forall n \in \mathcal{N}, \nonumber
		\end{flalign}
	\end{subequations}
	where ${\bf L}_0 \triangleq \{ {\bf w}_i[n], {\bf R}_{\rm s}^0[n], {\bf R}_{\rm s}^1[n], f_{{\rm l}, i}[n], f_{{\rm u}, i}[n], t_0[n],$ $t_1[n],$ ${\bf u}_i[n]  \}$. In \eqref{p0b} and \eqref{p0c_v1}, $f_{{\rm l}, \max}$ and $f_{{\rm u}, \max}$ denote the maximum CPU frequency for each UAV and the MEC server, respectively. In \eqref{p0c} and \eqref{p0d}, $P_{{\rm U}, \max}$ and $P_{{\rm AP}, \max}$ denote the transmit power budget for each UAV and the aggregate transmit power budget across all APs, respectively. In \eqref{p0i}, $V_{\max}$ denotes the maximum velocity of each UAV. In \eqref{p0j}, ${\bf u}_k^{\rm I}$ and ${\bf u}_k^{\rm F}$ denote the initial and final locations of the $k$-th UAV, respectively. In \eqref{p0k}, \eqref{p0l}, and \eqref{p0m}, $D_{\min}$ denotes the minimum distance between any two flying objects to avoid collision.

    \section{Joint Resource Allocation and UAV Trajectory Optimization}

    Problem ${\bf P}_0$ is hard to solve due to the non-convex objective function \eqref{p0a} and constraints \eqref{con_sen}, \eqref{con_com1}, \eqref{con_com2}, \eqref{con_covert}, \eqref{p0k}, \eqref{p0l}, and \eqref{p0m}. Moreover, the strong coupling among communication resources, sensing resources, computation resources, and UAV trajectories renders the considered problem more intractable. To address Problem ${\bf P}_0$, an AO-based framework is employed, which decomposes the original problem into two subproblems: 1) joint communication, sensing, and computation resources optimization subproblem, and 2) UAV trajectory optimization subproblem. The processes for handling each subproblem are given in the following subsections.

    \subsection{Joint Communication, Sensing, and Computation Resources Optimization}
    
    In this subsection, we jointly optimize the transmit beamformers $\{{\bf w}_k[n], {\bf R}_{\rm s}^0[n],  {\bf R}_{\rm s}^1[n]  \}$, computational resource allocation $\{f_{{\rm l}, k}[n], f_{{\rm u}, k}[n]\}$, and time allocation $\{t_0[n], t_1[n] \}$ with given UAV trajectories $\{ {\bf u}_k[n] \}$. It is observed that when $\{{\bf u}_k[n] \}$ is fixed, ${\bf L}_1[n] \triangleq  \{ {\bf w}_k[n], {\bf R}_{\rm s}^0[n], {\bf R}_{\rm s}^1[n],$$f_{{\rm l}, k}[n],$$ f_{{\rm u}, k}[n], t_0[n], t_1[n] \}$ are independent with ${\bf L}_1[n']$ for any $n'\neq n$. Thus, the resource allocation subproblem can be decomposed into $N$ independent {sub-subproblems} to be solved in parallel. Without loss of generality, we focus on the joint  resource allocation {sub-subproblem} of the $n$-th time slot, where the time slot index $n$ is omitted in this subsection for convenience. Specifically, the resource allocation {sub-subproblem} in the $n$-th time slot is given by
    \begin{subequations}
		\begin{flalign}
			{{\bf P}_1}: & \mathop{\min}  \limits_{ {\bf L}_1 }    ~E_{\rm C} + E_{\rm S}  + \sum\nolimits_{k \in \mathcal{K}} \left(  E_{{\rm l}, k} + E_{{\rm u}, k}    \right)
			  \label{p1a}  \\
			{\rm s.t.} ~
		& \eqref{con_sen}, \eqref{con_com1}, \eqref{con_com2}, \eqref{con_covert}, \eqref{p0b}, \eqref{p0c_v1}, \eqref{p0c}, \eqref{p0d}, \eqref{p0e}, 
        \nonumber \\
		& \eqref{p0f}, \eqref{p0g}, ~\forall {\bf q}_{\rm T} \in  \mathcal{Q},~\forall i, k \in \mathcal{K}, i \neq k,~\forall l \in \mathcal{L}. \nonumber
		\end{flalign}
	\end{subequations}
    Problem ${\bf P}_1$ is still non-convex due to constraints \eqref{con_sen}, \eqref{con_com1}, \eqref{con_com2}, \eqref{con_covert} and the objective function \eqref{p1a}.

    To handle the coupling among $t_0$, $t_1$, $f_{{\rm u}, k}$, and other optimization variables, the non-negative $t_0$, $t_1$, and $f_{{\rm u}, k}$ are substituted by $e^{\tau_0}$, $e^{\tau_1}$, and $e^{z_k}$, respectively.     The non-convex constraints \eqref{con_sen}, \eqref{con_com1}, and \eqref{con_com2} as well as \eqref{p0c_v1} and \eqref{p0e} are respectively re-expressed as
    \begin{flalign}
        &   \frac{e^{\tau_1} {\rm Tr}\left( {\bf G} {\bf R}_{\rm s}^1 {\bf G}^H  \right) }{ \sum_{k \in \mathcal{K}} {\rm Tr}\left({\bf H}_k {\bf w}_k {\bf w}_k^H {\bf H}_k^H   \right) + M N_{\rm R} \sigma_{\rm R}^2  } + \nonumber \\
        &~~~~~~~~~~~~~~~~~~~~~~~~~\frac{e^{\tau_o} {\rm Tr}\left( {\bf G} {\bf R}_{\rm s}^0 {\bf G}^H  \right)  }{M N_{\rm R} \sigma_{\rm R}^2} \geq \Gamma_{\min} \Delta_{\rm T}, \label{con_sen_r1}  \\
        & \frac{f_{{\rm l}, k}}{D_k} \Delta_{\rm T} + e^{\tau_1} B \log_2 \left( 1 + \Gamma_k \right) \geq I_k, \label{con_com1_r1} \\
        & f_{{\rm l}, k} \Delta_{\rm T} + e^{\tau_0+z_k} \geq I_k D_k \label{con_com2_r1},  \\
        & \sum\nolimits_{k \in \mathcal{K}} e^{z_k}  \leq f_{{\rm u}, \max}, \label{f_u_max}  \\
        & {\rm and}~e^{\tau_0} + e^{\tau_1} \leq \Delta_{\rm T}. \label{deltat}
    \end{flalign}

    However, the strong coupling in constraints \eqref{con_sen_r1} and \eqref{con_com1_r1}  still remains a crucial challenge. To handle \eqref{con_sen_r1}, auxiliary optimization variables $a_0$, $a_1$, and $b$ are introduced such that
    \begin{flalign}
        &{\rm Tr} \left({\bf G} {\bf R}_{\rm s}^0 {\bf G}^H  \right) \geq e^{a_0}, ~{\rm Tr}\left({\bf G}{\bf R}_{\rm s}^1 {\bf G}^H  \right) \geq e^{a_1},
        \label{a0a1} \\
        & \sum\nolimits_{k\in \mathcal{K}}   {\bf w}_k^H {\bf H}_k^H {\bf H}_k {\bf w}_k + M N_{\rm R} \sigma_{\rm R}^2 \leq e^b. \label{b}
    \end{flalign}
    Therefore, we can rewrite \eqref{con_sen_r1} as
    \begin{flalign}
           M N_{\rm R} \sigma_{\rm R}^2 e^{\tau_1+a_1-b}  + e^{\tau_o+a_0} \geq M N_{\rm R} \sigma_{\rm R}^2 \Gamma_{\min} \Delta_{\rm T}. \label{con_sen_r2}
    \end{flalign}
    As for \eqref{con_com1_r1}, we introduce auxiliary variables $r_k$, $\gamma_k$, and $\zeta_k$ satisfying that
    \begin{flalign}
        & \log_2 \left( 1 + e^{\gamma_k} \right) \geq e^{r_k}, \label{rk}
         \\ 
        &   {\bf w}_k^H {\bf H}_k^H {\bf H}_k {\bf w}_k          \geq e^{\gamma_k+\zeta_k},  \label{gammak}
        \\
        & \sum_{i \in \mathcal{K} \setminus \{k\} } {\bf w}_i^H {\bf H}_i^H {\bf H}_i{\bf w}_i  + {\rm Tr} \left( {\bf G} {\bf R}_{\rm s}^1 {\bf G}^H    \right) + M N_{\rm R} \sigma_{\rm R}^2  \leq e^{\zeta_k}, \label{zetak}
    \end{flalign}
    such that we reformulate \eqref{con_com1_r1}  as
    \begin{flalign}
        f_{{\rm l}, k} \Delta_{\rm T} + B D_k e^{\tau_1+r_k}  \geq I_k D_k. \label{con_com1_r2}
    \end{flalign}

    We transform the non-convex constraint \eqref{con_covert} to the following convex form:
    \begin{flalign}
        \sum\nolimits_{k \in \mathcal{K}} {\bf w}_k^H {\bf h}_{l, k} {\bf h}_{l, k}^H {\bf w}_k ~ +~& {\bf g}_l^H \left(  {\bf R}_{\rm s}^1 -  {\bf R}_{\rm s}^0  \right) {\bf g}_l  \nonumber \\
        &\leq  \mu_{\max} \left( {\bf g}_l^H {\bf R}_{\rm s}^0 {\bf g}_l + \sigma_l^2    \right). \label{con_covert_r1}
    \end{flalign}

With the definition of $\tau_0$, $\tau_1$, and $z_k$, the objective function \eqref{p1a} is rewritten as
    \begin{flalign}
        E_{\rm sum} \triangleq & ~e^{\tau_1}  \sum_{k\in\mathcal{K}} {\bf w}_k^H {\bf w}_k + e^{\tau_0} {\rm Tr}\left({\bf R}_{\rm s}^0\right) + e^{\tau_1} {\rm Tr}\left({\bf R}_{\rm s}^1  \right) + \nonumber  \\
        & 
        ~\sum_{k \in \mathcal{K}} \left( v_{\rm l} f_{{\rm l}, k}^3 \Delta_{\rm T}  + v_{\rm u} e^{\tau_0+3z_k}    \right),  \label{esum}
    \end{flalign}
    which involves coupling $\tau_o$, $\tau_1$, ${\bf w}_k$, ${\bf R}_{\rm s}^{0}$, and ${\bf R}_{\rm s}^1$. To tackle it, we introduce auxiliary variables $p_0$, $p_1$, and $p_2$ such that
    \begin{flalign}
        &{\rm Tr} \left( {\bf R}_{\rm s}^0  \right) \leq e^{p_0}, ~ {\rm Tr}\left( {\bf R}_{\rm s}^1 \right) \leq e^{p_1},  \label{p0p1} \\
        & \sum\nolimits_{k \in \mathcal{K}} {\bf w}_k^H {\bf w}_k \leq e^{p_2}.   \label{p2}
    \end{flalign}
    Then, $E_{\rm sum}$ in \eqref{esum} is expressed as a convex function $\widehat{E}_{\rm sum}$, which is given by
    \begin{flalign}
        & \widehat{E}_{\rm sum} = \\
        &~~ e^{\tau_1+p_2} + e^{\tau_0 + p_0} + e^{\tau_1+p_1} + \sum_{k \in \mathcal{K}} \left( v_{\rm l} f_{{\rm l}, k}^3 \Delta_{\rm T} + v_{\rm u} e^{\tau_0 +3 z_k}   \right). \nonumber
    \end{flalign}
    
    After the above transformations, Problem ${\bf P}_1$ can be equivalently reformulated as
    \begin{subequations}
	\begin{flalign}
			{{\bf P}_{1.1}}: & \mathop{\min}  \limits_{ {\bf L}_2 }    ~   \widehat{E}_{\rm sum}
			  \label{p1.1a}  \\
			{\rm s.t.} ~
		& \eqref{p0b}, \eqref{p0c},  \eqref{p0d}, \eqref{p0g}, \eqref{con_com2_r1}, \eqref{f_u_max},  \eqref{deltat}, \eqref{a0a1}, \eqref{b}, \nonumber \\
        &\eqref{con_sen_r2}, \eqref{rk}, \eqref{gammak}, \eqref{zetak}, \eqref{con_com1_r2}, \eqref{con_covert_r1}, \eqref{p0p1}, \eqref{p2},
        \nonumber \\
		& \forall {\bf q}_{\rm T} \in  \mathcal{Q},~\forall i, k \in \mathcal{K}, i \neq k,~\forall l \in \mathcal{L}, \nonumber
        \end{flalign}
    \end{subequations}
    where ${\bf L}_2 \triangleq$ $$ \{ {\bf w}_k, {\bf R}_{\rm s}^0, {\bf R}_{\rm s}^1,f_{{\rm l}, k}, \tau_0, \tau_1, z_k, a_0, a_1, b, r_k, \gamma_k,\zeta_k, p_0, p_1, p_2 \}.$$
    Problem ${\bf P}_{1.1}$ is still non-convex due to constraints \eqref{con_com2_r1}, \eqref{b}, \eqref{con_sen_r2}, \eqref{rk}, \eqref{gammak}, \eqref{zetak}, \eqref{con_com1_r2}, \eqref{p0p1}, and \eqref{p2}. Nevertheless, as both sides of these constraints are convex functions, they can be approximated by utilizing first-order Taylor expansion based on a given feasible point $\widetilde{\bf L}_2 \triangleq$ 
    $$ \{ \widetilde{\bf w}_k, \widetilde{\bf R}_{\rm s}^0, \widetilde{\bf R}_{\rm s}^1,\widetilde{f}_{{\rm l}, k}, \widetilde{\tau}_0, \widetilde{\tau}_1, \widetilde{z}_k, \widetilde{a}_0, \widetilde{a}_1, \widetilde{b}, \widetilde{r}_k, \widetilde{\gamma}_k,\widetilde{\zeta}_k, \widetilde{p}_0, \widetilde{p}_1, \widetilde{p}_2 \}.$$
    Specifically, constraints \eqref{con_com2_r1}, \eqref{b}, \eqref{con_sen_r2}, \eqref{rk}, \eqref{gammak}, \eqref{zetak}, \eqref{con_com1_r2}, \eqref{p0p1}, and \eqref{p2} are respectively approximated by
    \begin{flalign}
        & f_{{\rm l}, k} \Delta_{\rm T} + \kappa(\tau_0+z_k, \widetilde{\tau}_0 + \widetilde{z}_k ) \geq I_k D_k, \label{tilde_tau_z} \\
        & \sum\nolimits_{k \in \mathcal{K}} {\bf w}_k^H {\bf H}_k^H {\bf H}_k {\bf w}_k + M N_{\rm R} \sigma_{\rm R}^2 \leq \kappa(b, \widetilde{b}), \label{tilde_b} \\
        & M N_{\rm R} \sigma_{\rm R}^2 \kappa(\tau_1+a_1-b, \widetilde{\tau}_1 + \widetilde{a}_1 - \widetilde{b}) + \kappa(\tau_0 + a_0, \widetilde{\tau}_0 + \widetilde{a}_0  )  \nonumber \\
        & ~~~~~~~~~~~~~~~~~~~~~~~~~~~~~~~~~~~~~~ \geq M N_{\rm R} \sigma_{\rm R}^2 \Gamma_{\min} \Delta_{\rm T},  \label{tilde_tau_a_b}  \\
        & \log_2\left( 1 + e^{\widetilde{\gamma}_k}  \right) + \frac{e^{\widetilde{\gamma}_k}  \left( {\gamma}_k - \widetilde{\gamma}_k \right) }{\ln2 \left( 1 + e^{\widetilde{\gamma}_k}  \right) } \geq e^{r_k},  \label{tilde_gamma}  \\
        & \widetilde{\bf w}_k^H {\bf H}_k^H {\bf H}_k \widetilde{\bf w}_k + 2 {\rm Re} \left( \widetilde{\bf w}_k^H {\bf H}_k^H {\bf H}_k \left( {\bf w}_k - \widetilde{\bf w}_k \right) \right) \geq e^{\gamma_k + \zeta_k},   \label{tilde_w} \\
        & \sum_{i \in \mathcal{K} \setminus \{k\}} {\bf w}_i^H {\bf H}_i^H {\bf H}_i {\bf w}_i + {\rm Tr} \left({\bf G} {\bf R}_{\rm s}^1 {\bf G}^H \right) + M N_{\rm R} \sigma_{\rm R}^2 \nonumber \\
        &~~~~~~~~~~~~~~~~~~~~~~~~~~~~~~~~~~~~~~~~~~~~~~
        \leq \kappa(\zeta_k, \widetilde{\zeta}_k), \label{tilde_zeta} \\
        & f_{{\rm l}, k} \Delta_{\rm T} + B D_k \kappa(\tau_1+r_k, \widetilde{\tau}_1 + \widetilde{r}_k  ) \geq I_k D_k, \label{tilde_tau_r}  \\
        & {\rm Tr} \left( {\bf R}_{\rm s}^0 \right) \leq \kappa(p_0, \widetilde{p}_0),~{\rm Tr} \left( {\bf R}_{\rm s}^1 \right) \leq \kappa(p_1, \widetilde{p}_1), \label{tilde_p0p1} \\
        & \sum\nolimits_{k\in \mathcal{K}} {\bf w}_k^H {\bf w}_k  \leq \kappa(p_2, \widetilde{p}_2), \label{tilde_p2}
        \end{flalign}
    where $\kappa(x, \widetilde{x}) \triangleq e^{\widetilde{x}} ( 1+x - \widetilde{x} ) $.

     Problem ${\bf P}_{1.1}$ can be approximated by the following convex optimization problem:
     \begin{subequations}
	\begin{flalign}
			{{\bf P}_{1.2}}: & \mathop{\min}  \limits_{ {\bf L}_2 }    ~   \widehat{E}_{\rm sum}
			  \label{p1.2a}  \\
			{\rm s.t.} ~
		& \eqref{p0b}, \eqref{p0c},  \eqref{p0d}, \eqref{p0g}, \eqref{f_u_max},  \eqref{deltat}, \eqref{a0a1}, \eqref{con_covert_r1}, \eqref{tilde_tau_z}, \nonumber \\
        & \eqref{tilde_b}, \eqref{tilde_tau_a_b}, \eqref{tilde_gamma}, \eqref{tilde_w}, \eqref{tilde_zeta}, \eqref{tilde_tau_r}, \eqref{tilde_p0p1}, \eqref{tilde_p2},
        \nonumber \\
		& \forall {\bf q}_{\rm T} \in  \mathcal{Q},~\forall i, k \in \mathcal{K}, i \neq k,~\forall l \in \mathcal{L}. \nonumber
        \end{flalign}
    \end{subequations}
    In order to improve the approximation precision, we employ the SCA method by iteratively updating $\widetilde{\bf L}_2$. The proposed SCA-based algorithm for addressing Problem ${\bf P}_1$ is summarized in Algorithm \ref{alg:1}.
   \begin{algorithm}
    \caption{The Proposed SCA-based Algorithm for Solving Problem {${\bf P}_1$}.} \label{alg:1}
\begin{algorithmic}
  \State{\textbf{Initialization}:}
  \State{~ Obtain a feasible $\widetilde{\bf L}_2$ to Problem ${\bf P}_{1.1}$;}
  \While{\rm the stop criterion is not satisfied}
      \State{Obtain the optimal ${\bf L}_2^{\star}$ via solving Problem ${\bf P}_{1.2}$ with given $\widetilde{\bf L}_2$ and UAV locations $\{ {\bf u}_k \}$;}
      \State{Update $\widetilde{\bf L}_2 := {\bf L}_2^{\star}$;}
  \EndWhile
  \State{~ Obtain $t_0^{\star} = e^{\tau_0^{\star}}$, $t_1^{\star} =  e^{\tau_1^{\star}}$, $f_{{\rm u}, k}^{\star} = e^{z_k^{\star}}$;}
  \State{\textbf{Return}: ${\bf L}_1^{\star} \triangleq \{ {\bf w}_k^{\star}, {({\bf R}_{\rm s}^0)}^{\star}, {({\bf R}_{\rm s}^1)}^{\star}, f_{{\rm l}, k}^{\star}, f_{{\rm u}, k}^{\star}, t_0^{\star}, t_1^{\star} \}$.}
\end{algorithmic}
\end{algorithm}

    \subsection{UAV Trajectory Optimization}
    In this subsection, we optimize the UAV trajectories $\{{\bf u}_k[n]\}$ with pre-defined communication, sensing, and computation resource allocation scheme, i.e., $\{ {\bf L}_1[n] \}$. The corresponding optimization problem is given by
    \begin{subequations}
		\begin{flalign}
			{{\bf P}_2}: & \mathop{\min}  \limits_{ \{ {\bf u}_k[n] \} }   \sum\nolimits_{n \in \mathcal{N}}   \sum\nolimits_{k \in \mathcal{K}}  P(||{\bf v}_k [n]||)  \Delta_{\rm T}  
			  \label{p2a}  \\
			{\rm s.t.} ~
		& \eqref{con_sen}, \eqref{con_com1}, \eqref{con_covert}, \eqref{p0i}, \eqref{p0j}, \eqref{p0k}, \eqref{p0l}, \eqref{p0m},    \nonumber \\
		& \forall {\bf q}_{\rm T}[n] \in  \mathcal{Q},~\forall i, k \in \mathcal{K}, i \neq k,~\forall l \in \mathcal{L},~\forall n \in \mathcal{N}. \nonumber
		\end{flalign}
	\end{subequations}
Problem ${\bf P}_2$ is non-convex due to the objective function \eqref{p2a} and constraints \eqref{con_sen}, \eqref{con_com1}, \eqref{con_covert}, \eqref{p0k}, \eqref{p0l}, and \eqref{p0m}. 

Let us handle the objective function \eqref{p2a} first. By introducing auxiliary optimization variables $v_{1, k}[n]$ and $v_{2, k}[n]$ which satisfy that 
\begin{flalign}
    & v_{1, k}[n] \Delta_{\rm T} \geq || {\bf u}_k[n+1] - {\bf u}_k[n]  ||, \label{v1k} \\
    & v_{2, k}^2[n] + \frac{ ||{\bf u}_k[n+1] - {\bf u}_k[n] ||^2     }{v_0^2 \Delta_{\rm T}^2}  \geq \frac{1}{ v_{2, k}^2[n] }, \label{v2k}
\end{flalign}
\eqref{p2a} is reformulated as the following convex function:
\begin{flalign}
    & E_{\rm fly} \left( v_{1, k}[n], v_{2, k}[n]    \right) \triangleq      \\
    & \left(P_0 \left(  1 + \frac{3 v_{1, k}^2[n]  }{U_{\rm tip}^2}    \right) + \frac{1}{2} d_0 \rho_0 s A  v_{1,k}^3[n]  + P_{\rm H} v_{2, k}[n]   \right) \Delta_{\rm T}. \nonumber
\end{flalign}
As for constraints \eqref{con_sen}, \eqref{con_com1}, and \eqref{con_covert}, they are respectively simplified as
    \begin{flalign}
    &
    \underbrace{t_0[n] {\rm Tr} \left( {\bf G}[n] {\bf R}_{\rm s}^0[n] {\bf G}^H[n]  \right)  + t_1[n] {\rm Tr}\left( {\bf G}[n] {\bf R}_{\rm s}^1[n] {\bf G}^H[n]   \right)}_{\triangleq G_1({\bf q}_{\rm T}[n])} 
      \nonumber \\
    &~~~ - M N_{\rm R} \sigma_{\rm R}^2 \Gamma_{\min} \Delta_{\rm T} \geq \sum_{k \in \mathcal{K}} \Psi_k({\bf u}_k[n])  \times
    \nonumber \\
    &\underbrace{\left( \Gamma_{\min} \Delta_{\rm T} -  \frac{ t_0[n] {\rm Tr} \left( {\bf G}[n] {\bf R}_{\rm s}^0[n] {\bf G}^H[n]  \right)   }{M N_{\rm R} \sigma_{\rm R}^2  }  \right)}_{\triangleq G_2({\bf q}_{\rm T}[n])  }, 
    \label{con_sen_tra} \\
    & \Psi_k({\bf u}_k[n])        \geq \underbrace{\left(2^{ \frac{I_k[n] D_k - f_{{\rm l}, k}[n] \Delta_{\rm T}} {t_1[n] B D_k }}  -1 \right)}_{\triangleq R_k[n] } \Bigg( \sum_{i \in \mathcal{K} \setminus \{k\} } \Psi_i( {\bf u}_i[n])    \nonumber \\
    &+ {\rm Tr} \left( {\bf G}[n] {\bf R}_{\rm s}^1[n] {\bf G}^H[n]    \right) + M N_{\rm R} \sigma_{\rm R}^2     \Bigg),  \label{con_com_tra} \\
    & \sum_{k \in \mathcal{K}} \Omega_{l, k}({\bf u}_k[n]) + {\bf g}_l^H \left(  {\bf R}_{\rm s}^1[n] -  {\bf R}_{\rm s}^0[n]  \right) {\bf g}_l \leq \nonumber \\
    &~~~~~~~~~~~~~~~~~~~~~~~~~~~~~~~\mu_{\max} \left(  {\bf g}_l^H {\bf R}_{\rm s}^0[n] {\bf g}_l + \sigma_l^2  \right), \label{con_covert_tra}
    \end{flalign}
   where
   \begin{flalign}
       & \Psi_k({\bf u}_k[n]) = {\rm Tr} \left( {\bf H}_k[n] {\bf w}_k[n] {\bf w}_k^H[n] {\bf H}_k^H[n]  \right), \\
       & \Omega_{l, k}({\bf u}_k[n]) = {\bf h}_{l, k}^H[n] {\bf w}_k[n] {\bf w}_k^H[n] {\bf h}_{l, k}[n] . 
   \end{flalign}

Then, Problem ${\bf P}_2$ is reformulated as
  \begin{subequations}
		\begin{flalign}
			{{\bf P}_{2.1}}: & \mathop{\min}  \limits_{ \{ {\bf u}_k[n], v_{1, k}[n], v_{2, k}[n] \} }   \sum\limits_{n \in \mathcal{N}}   \sum\limits_{k \in \mathcal{K}}  E_{\rm fly} \left( v_{1, k}[n], v_{2, k}[n]    \right)  
			  \label{p2.1a}  \\
			{\rm s.t.} ~
		&  \eqref{p0i}, \eqref{p0j}, \eqref{p0k}, \eqref{p0l}, \eqref{p0m}, \eqref{v1k}, \eqref{v2k}, \eqref{con_sen_tra}, \eqref{con_com_tra},    \nonumber \\
		&\eqref{con_covert_tra},~ \forall {\bf q}_{\rm T}[n] \in  \mathcal{Q},~\forall i, k \in \mathcal{K}, i \neq k,~\forall l \in \mathcal{L},~\forall n \in \mathcal{N}. \nonumber
		\end{flalign}
    \end{subequations}
Problem ${\bf P}_{2.1}$ is non-convex due to constraints \eqref{p0k}, \eqref{p0l}, \eqref{p0m}, \eqref{v2k}, \eqref{con_sen_tra}, \eqref{con_com_tra}, and \eqref{con_covert_tra}. Furthermore, the UAV trajectories ${\bf u}_k[n]$ simultaneously affect both path-loss terms and steering vectors in ${\Psi}_k({\bf u}_k[n])$ and ${\Omega}_{l,k}({\bf u}_k[n])$, introducing additional non-convexity which exacerbates the intractability of the considered problem. 

To handle the non-convex constraints in Problem ${\bf P}_{2.1}$, a trust-region-based algorithm \cite{trust_region} is proposed, which is implemented in an iterative manner. Denote the local trajectory point in the $\iota$-th iteration by $\{ \widetilde{\bf u}_k^{\iota}[n]\}$, and the corresponding auxiliary optimization variables $\widetilde{v}_{1, k}^{\iota}[n]$ and $\widetilde{v}_{2, k}^{\iota}[n]$ as
\begin{flalign}
    &  \widetilde{v}_{1, k}^{\iota}[n] \triangleq \frac{ ||\widetilde{\bf u}_k^{\iota}[n+1] - \widetilde{\bf u}_k^{\iota}[n]||  }{\Delta_{\rm T}}, \label{iota_v1} \\
    & \widetilde{v}_{2, k}^{\iota}[n] \triangleq \sqrt{\left(\sqrt{1 + \frac{ \left(  \widetilde{v}^{\iota}_{1, k}[n] \right)^4  }{4 v_0^4}} - \frac{\left(  \widetilde{v}_{1, k}^{\iota}[n]   \right)^2}{2 v_0^2}  \right)}.
    \label{iota_v2}
\end{flalign}
The detailed processes of non-convex constraints \eqref{p0k}, \eqref{p0l}, \eqref{p0m}, \eqref{v2k}, \eqref{con_sen_tra}, \eqref{con_com_tra}, and \eqref{con_covert_tra} are given in the following.

First, we respectively relax \eqref{p0k}, \eqref{p0l}, \eqref{p0m}, and \eqref{v2k}, whose both sides are convex functions, as 
\begin{flalign}
    & 2 (\widetilde{\bf u}^{\iota}_k[n] - \widetilde{\bf u}^{\iota}_i[n])^T  ({\bf u}_k[n] -{\bf u}_i[n] )  - || \widetilde{\bf u}^{\iota}_k[n] -  \widetilde{\bf u}^{\iota}_i[n]  ||^2  + \nonumber \\
    &~~~~~~~~~~~~~~~~~~~~~~~~~~~~~~~
    (H_k-H_i)^2 \geq D_{\min}^2, \label{tilde_uk1} \\
    & 2 (\widetilde{\bf q}_k^{\iota}[n] - {\bf q}_l^{\rm W})^T ({\bf q}_k^{\rm U}[n] - {\bf q}_l^{\rm W} )  -  \nonumber \\
    &~~~~~~~~~~~~~~~~~~~~~~~~~~~~~~~
    || \widetilde{\bf q}_k^{\iota}[n] - {\bf q}_l^{\rm W}||^2
     \geq D_{\min}^2,  \label{tilde_uk2} \\
    & 2 (\widetilde{\bf q}_k^{\iota}[n] - {\bf q}_{\rm T}[n])^T ({\bf q}_k^{\rm U}[n] - {\bf q}_{\rm T}[n] ) - \nonumber \\
    &~~~~~~~~~~~~~~~~~~~~~~~~~~~~~~~
    || \widetilde{\bf q}_k^{\iota}[n] - {\bf q}_{\rm T}[n]  ||^2 \geq D_{\min}^2,   \label{tilde_uk3} \\
    & {\rm and}~ 2 \widetilde{v}^{\iota}_{2, k}[n] v_{2, k}[n] -  
    \left(\widetilde{v}^{\iota}_{2, k}[n]\right)^2 
     -  \frac{|| \widetilde{\bf u}^{\iota}_k[n+1] - \widetilde{\bf u}^{\iota}_k[n] ||^2}{v_0^2 \Delta_{\rm T}^2} +
       \nonumber   \\
    & \frac{ 2(\widetilde{\bf u}^{\iota}_k[n+1] - \widetilde{\bf u}^{\iota}_k[n])^T ( {\bf u}_k[n+1] - {\bf u}_k[n] )  }{v_0^2 \Delta_{\rm T}^2}
        \geq \frac{1}{ v_{2, k}^2[n]  },   \label{tilde_v2k}
\end{flalign}
where $\widetilde{\bf q}_k^{\iota}[n] \triangleq [(\widetilde{\bf u}^{\iota}_k[n])^T, H_k]^T$.

Then, we handle \eqref{con_sen_tra}, \eqref{con_com_tra}, and \eqref{con_covert_tra}, whose non-convexity is caused by ${\Psi}_k({\bf u}_k[n])$ and ${\Omega}_{l, k}({\bf u}_k[n])$. Similar with \cite{trust_region_appro}, ${\Psi}_k({\bf u}_k[n])$ and ${\Omega}_{l, k}({\bf u}_k[n])$ are approximated by their first-order Taylor expansions, i.e., 
\begin{flalign}
    & \Psi_k({\bf u}_k[n]) \approx \widetilde{\Psi}_k({\bf u}_k[n], \widetilde{\bf u}_k^{\iota}[n]) \triangleq  
    \label{psi_tilde}   \\
    & \Psi_k(\widetilde{\bf u}^{\iota}_k[n]) + \left(\left. \frac{{\rm d} {\Psi}_k ({\bf u}_k[n]) }{{\rm d}{\bf u}_k[n] } \right|_{ \widetilde{\bf u}_k^{\iota}[n]} \right)^T \left( {\bf u}_k[n] - \widetilde{\bf u}_k^{\iota}[n]    \right),  \nonumber \\
    & \Omega_{l, k} ({\bf u}_k[n]) \approx \widetilde{\Omega}_{l, k} ({\bf u}_k[n], \widetilde{\bf u}_k^{\iota}[n]  ) \triangleq
    \label{omega_tilde}    \\
    & \Omega_{l, k}(\widetilde{\bf u}^{\iota}_k[n]  ) +
    \left( \left.  \frac{{\rm d} \Omega_{l, k}({\bf u}_k[n])  }{{\rm d} {\bf u}_k[n]} \right|_{ \widetilde{\bf u}^{\iota}_k[n]  }  \right)^T \left( {\bf u}_k[n] - \widetilde{\bf u}^{\iota}_k[n]    \right), \nonumber
\end{flalign}
where ${\rm d}\Psi_k({\bf u}_k[n]) / {\rm d} {\bf u}_k[n]$ and ${\rm d} \Omega_{l, k} ({\bf u}_k[n]) / {\rm d} {\bf u}_k[n]$ denote the derivative of $\Psi_k({\bf u}_k[n])$ and $\Omega_{l, k}({\bf u}_k[n])$ w.r.t. ${\bf u}_k[n]$, respectively. The corresponding calculation processes are given in Appendix A. Based on  \eqref{psi_tilde} and \eqref{omega_tilde}, we reach approximated forms of \eqref{con_sen_tra}, \eqref{con_com_tra}, and \eqref{con_covert_tra} as
\begin{flalign}
    &  G_1({\bf q}_{\rm T}[n] ) -  M N_{\rm R} \sigma_{\rm R}^2 \Gamma_{\min} \Delta_{\rm T} \geq \nonumber \\
    &~~~~~~~~~~~~~~~~~~~~~~~~~
   G_2({\bf q}_{\rm T}[n] )\sum_{k \in \mathcal{K}} \widetilde{\Psi}_k({\bf u}_k[n], \widetilde{\bf u}_k^{\iota}[n]),     \label{tilde_con_sen_tra} \\
    & \widetilde{\Psi}_k({\bf u}_k[n], \widetilde{\bf u}^{\iota}_k[n])        \geq R_k[n] \Bigg( \sum_{i \in \mathcal{K} \setminus \{k\} } \widetilde{\Psi}_i( {\bf u}_i[n], \widetilde{\bf u}^{\iota}_i[n])  +  \nonumber \\
    &~~~ {\rm Tr} \left( {\bf G}[n] {\bf R}_{\rm s}^1[n] {\bf G}^H[n]    \right) + M N_{\rm R} \sigma_{\rm R}^2     \Bigg),  \label{tilde_con_com_tra} \\
    & \sum_{k \in \mathcal{K}} \widetilde{\Omega}_{l, k}({\bf u}_k[n], \widetilde{\bf u}_k^{\iota}[n]) + {\bf g}_l^H \left(  {\bf R}_{\rm s}^1[n] -  {\bf R}_{\rm s}^0[n]  \right) {\bf g}_l \leq \nonumber \\
    &~~~~~~~~~~~~~~~~~~~~~~~~~~~~~~~\mu_{\max} \left(  {\bf g}_l^H {\bf R}_{\rm s}^0[n] {\bf g}_l + \sigma_l^2  \right). \label{tilde_con_covert_tra}
\end{flalign}

Furthermore, in the $\iota$-th iteration, we impose the following constraint to control the approximation error in \eqref{psi_tilde} and \eqref{omega_tilde}:
\begin{flalign}
    || {\bf u}_k[n] - \widetilde{\bf u}_k^{\iota}[n] || \leq \omega^{\iota}, \label{trust_region}
\end{flalign}
where $\omega^{\iota}$ denotes the trust region radius in the $\iota$-th iteration. It is noted that when $\omega^{\iota}$ is chosen to be sufficiently small, the convergence of the proposed trust-region-based algorithm can be guaranteed \cite{trust_region_con}.

At last, Problem ${\bf P}_{2.1}$ is approximated by the following convex optimization problem:
\begin{subequations}
\begin{flalign}
	{{\bf P}_{2.2}}: & \mathop{\min}  \limits_{ \{ {\bf u}_k[n], v_{1, k}[n], v_{2, k}[n] \} }   \sum\limits_{n \in \mathcal{N}}   \sum\limits_{k \in \mathcal{K}}  E_{\rm fly} \left( v_{1, k}[n], v_{2, k}[n]    \right)  
			  \label{p2.2a}  \\
			{\rm s.t.} ~
		&  \eqref{p0i}, \eqref{p0j}, \eqref{v1k}, \eqref{tilde_uk1}, \eqref{tilde_uk2}, \eqref{tilde_uk3}, \eqref{tilde_v2k}, \eqref{tilde_con_sen_tra}, \eqref{tilde_con_com_tra}, \eqref{tilde_con_covert_tra},    \nonumber \\
		&\eqref{trust_region}, ~ \forall {\bf q}_{\rm T}[n] \in  \mathcal{Q},~\forall i, k \in \mathcal{K}, i \neq k,~\forall l \in \mathcal{L},~\forall n \in \mathcal{N}. \nonumber
\end{flalign}
\end{subequations}

The proposed iterative trust-region-based algorithm is summarized in Algorithm \ref{alg: trust}. Notably, in the $\iota$-th iteration, if the objective value of Problem ${\bf P}_{2.1}$ is not decreased compared to that in the $(\iota-1)$-th iteration, the trust region radius is reduced by $\omega^{\iota} := \omega^{\iota}/2$ and we resolve Problem ${\bf P}_{2.2}$. When $\omega^{\iota}$ is less than a pre-defined threshold $\omega_{\min}$, the iteration terminates. 

\begin{algorithm}
    \caption{The Proposed Trust-Region-based Algorithm for Solving Problem {${\bf P}_2$}} \label{alg: trust}
\begin{algorithmic}
  \State{\textbf{Initialization}:}
  \State{ ~ Set iteration index $\iota = 1$;}
  \State{ ~ Obtain a feasible $\{ \widetilde{\bf u}_k^{\iota}[n] \}$ to Problem ${\bf P}_{2}$;}
  \State{~ Obtain $\{ \widetilde{v}_{1, k}^{\iota}[n], \widetilde{v}_{2, k}^{\iota}[n] \}$ via \eqref{iota_v1} and \eqref{iota_v2} with given $\{ \widetilde{\bf u}_k^{\iota}[n] \}$, respectively; }
  \While{$\omega^{\iota} \geq \omega_{\min} $ }
      \State{Obtain the optimal $\{{\bf u}_k^{\star}[n], v_{1, k}^{\star}[n], v_{2, k}^{\star}[n]\}$ to Problem ${\bf P}_{2.2}$ with given $\{\widetilde{\bf u}_k^{\iota}[n],  \widetilde{v}_{1, k}^{\iota}[n], \widetilde{v}_{2, k}^{\iota}[n]   \}$;} 
      \If{\rm  the objective value in \eqref{p2.1a} decreases}
      \State{Update $\iota := \iota+1$ and $\{ \widetilde{\bf u}_k^{\iota}[n]\} := \{ {\bf u}_k^{\star}[n] \}$;}
      \State{Update $\{ \widetilde{v}_{1, k}^{\iota}[n], \widetilde{v}_{2, k}^{\iota}[n] \}$ via \eqref{iota_v1} and \eqref{iota_v2}; }
      \Else
      \State{ Update $\omega^{\iota} := \omega^{\iota}/2$;  }
      \EndIf
  \EndWhile
  \State{\textbf{Return}: $\{ {\bf u}_k^{\star}[n] \}$.}
\end{algorithmic}
\end{algorithm}

\section{Simulation Results}
This section presents the simulation results to evaluate our proposed algorithm as well as to reveal systematic insights for communication, sensing, and computation. Referring to \cite{bc_LAE_ISAC, rw_ISAC6, rw_covert5}, we consider $M=3$ ISAC APs, $K=2$ UAVs, and $L=2$ wardens, which are located in a square area of  $300~{\rm m} \times 300~{\rm m}$. The size of sensing area is $20~{\rm m} \times 20~{\rm m} \times 10~{\rm m}$ with vertical coordinates spanning the interval $[10~{\rm m}, 20~{\rm m}]$. The flight altitude of each UAV is set as $H_k = 100~{\rm m}, \forall k \in \mathcal{K}$, while that of each warden is set as $H_l^{\rm W} = 105~{\rm m}, \forall l \in \mathcal{L}$. For clarity, the remaining system parameters are specified in Table \ref{tab:parameter}.

\begin{table}[ht]
    \centering
    \caption{Simulation Parameters}
    \setlength{\tabcolsep}{4pt} 
    \begin{tabular}{c c||c c || c c}
        \hline
        \textbf{Notation} & \textbf{Value} & \textbf{Notation} & \textbf{Value} & \textbf{Notation} & \textbf{Value} \\ 
        \hline
        \hline
        $N_{\rm T}$ & 16 & $T$ & $30$ ${\rm s}$ & $P_0$ & $79.86$ ${\rm W}$  \\ 
        $N_{\rm R}$ & 2 & $N$ & $30$  & $P_{\rm H}$ & $88.63$ ${\rm W}$   
        \\ 
        $N_{\rm U}$ & 2 & $V_{\max}$ & $20$ ${\rm m/s}$ & $U_{\rm tip}$ & $120$ ${\rm m/s}$ 
        \\ 
        $C_0$ & $10^{-3}$  &  $v_0$  &  $4.03$ ${\rm m/s}$ & $d_0$ & $0.6$
        \\ 
        $B$ & $30$ ${\rm MHz}$   &  $v_{\rm l}$  &  $10^{-26}$   & $\rho_0$ & $1.225$ 
        \\  
        $I_k[n]$  &  $7$ ${\rm Mbit}$   & $v_{\rm u}$  &  $10^{-28}$  & s & $0.05$
        \\ [1.5pt]
        $\sigma_{\rm R}^2$ & $-100$ ${\rm dBW}$   & $D_k$ & $10^3$  &  $A$ & $0.503$ ${\rm m^2}$
        \\ [1.5pt]
        $\sigma_{l}^2$ & $-100$ ${\rm dBW}$ & $f_{{\rm l, max}}$ & $5$ ${\rm GHz}$  & $D_{\min}$ & $20$ ${\rm m}$
        \\
        $P_{\rm U, max}$ & $10$ ${\rm mW}$ & $f_{\rm u, max}$ & $50$ ${\rm GHz}$  & $\mu_{\max}$  & 0.0276 \\
        $P_{\rm AP, max}$ & $30$ ${\rm W}$ & $\Gamma_{\min}$ & 0.1 & $Q$ & 18
        \\ [0.5pt] \hline
    \end{tabular}
    \label{tab:parameter}
\end{table}

\begin{figure*}[!t]
		\centering
		\subfloat[Case 1: Trajectories with ${\bf q}_1^{\rm W}= (220, 30, 105)^T{\rm m}$ and ${\bf q}_2^{\rm W}= (220, 270, 105)^T{\rm m}$.
		 ]{\includegraphics[width=0.32\textwidth]{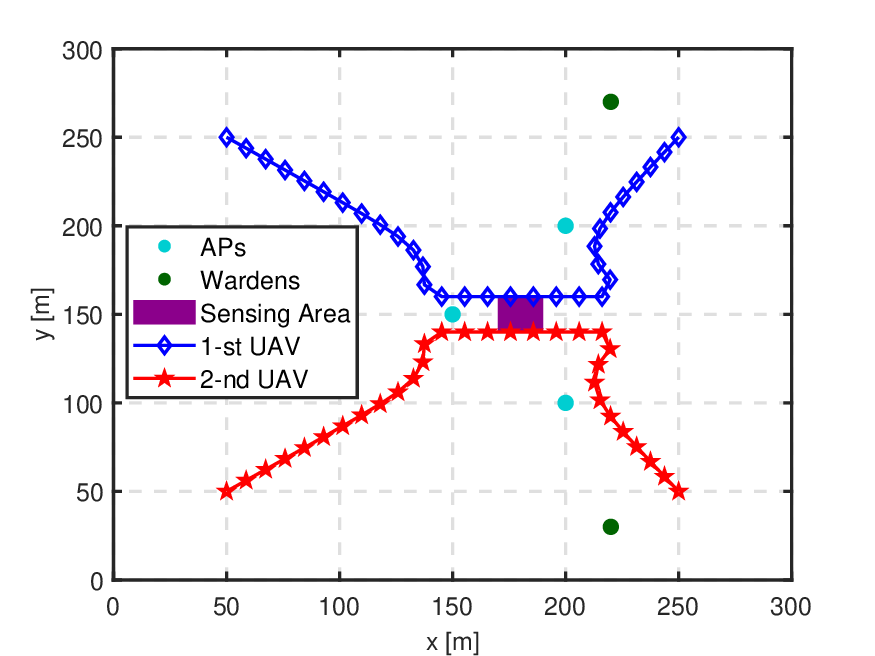}}
		\hfill
		\subfloat[Case 2: Trajectories with ${\bf q}_1^{\rm W}= (100, 80, 105)^T{\rm m}$ and ${\bf q}_2^{\rm W}= (100, 220, 105)^T{\rm m}$.]{\includegraphics[width=0.32\textwidth]{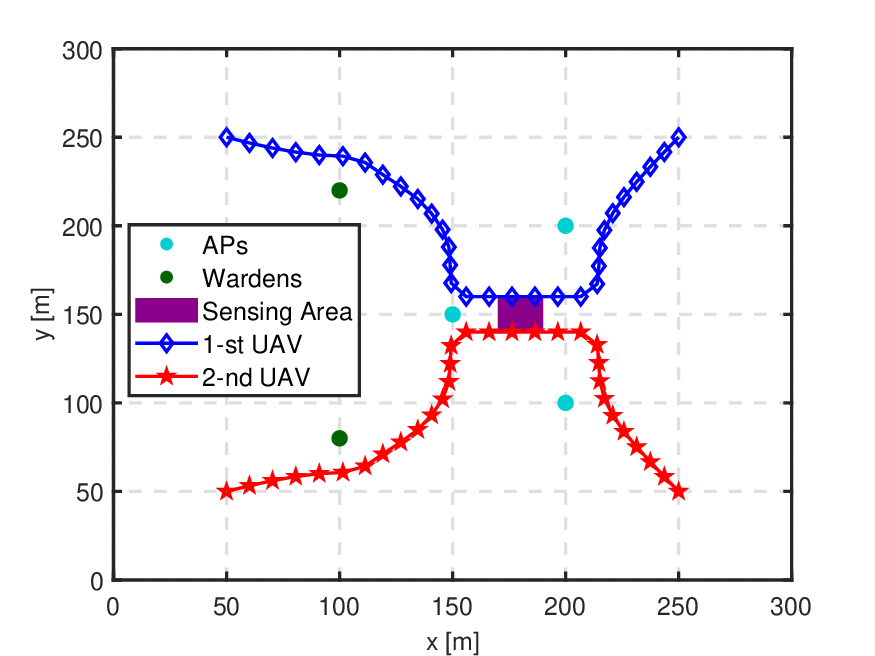}}
		\hfill
		\subfloat[Case 3: Trajectories with ${\bf q}_1^{\rm W}= (220, 130, 105)^T{\rm m}$ and ${\bf q}_2^{\rm W}= (220, 170, 105)^T{\rm m}$.]{\includegraphics[width=0.32\textwidth]{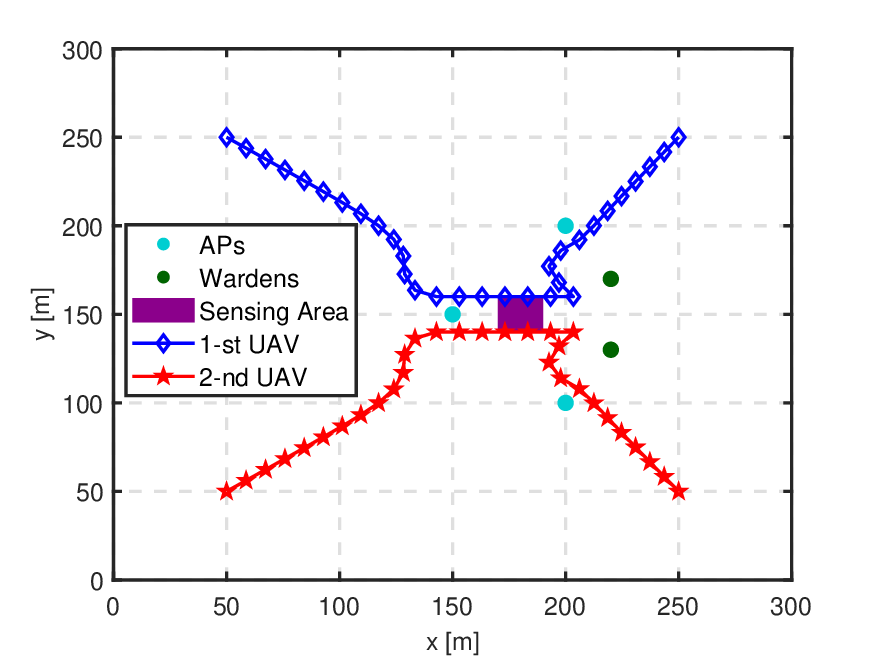}}
		\caption{Trajectories of UAVs under various warden locations.}
		\label{fig: tras}
	\end{figure*}

    Fig. \ref{fig: tras} illustrates the trajectories of UAVs under three cases of warden locations: Case 1 with ${\bf q}_1^{\rm W}=(220, 30, 105)^T~{\rm m}$ and ${\bf q}_2^{\rm W}=(220, 270, 105)^T~{\rm m}$; Case 2 with ${\bf q}_1^{\rm W}=(100, 80, 105)^T~{\rm m}$ and ${\bf q}_2^{\rm W}=(100, 220, 105)^T~{\rm m}$; and Case 3 with ${\bf q}_1^{\rm W}=(220, 130, 105)^T~{\rm m}$ and ${\bf q}_2^{\rm W}=(220, 170, 105)^T~{\rm m}$. As observed, in all cases, UAVs intend to approach APs to reduce the propagation loss of task offloading, thereby enabling more task bits to be transmitted efficiently. The reason is that handling tasks at the edge is far more cost-effective than onboard UAVs.  Besides, the three cases indicate that the UAVs actively evade wardens to prevent both physical collisions and information leakage. However, this evasive behavior may incur increased propulsion energy consumption and degrade the quality of offloading links.

    \begin{figure}[!t]
		\centering
		\includegraphics[width=0.48\textwidth]{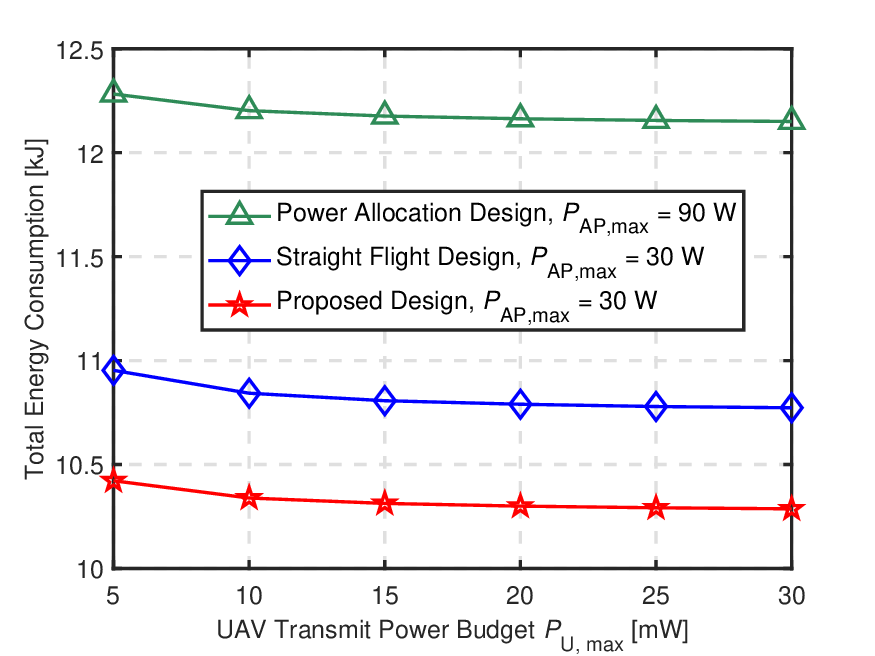}
		\caption{Comparison among the proposed, power allocation, and straight flight designs.}
		\label{fig: pumax}
    \end{figure}

    Fig. \ref{fig: pumax} shows the total energy consumption versus the UAV transmit power budget $P_{\rm U, max}$. For comparison, two benchmarks are included: the power allocation design and the straight flight design. The power allocation design adopts maximum ratio transmission for UAV task offloading and isotropic transmission for AP-based target sensing, while the straight flight design has each UAV  follow a linear trajectory at constant velocity between its initial and final locations. The AP transmit power budget is set by $P_{\rm AP, max} = 30~{\rm W}$ for both the proposed design and the straight flight design, while for the power allocation design, we increase the power budget to $P_{\rm AP, max} = 90~{\rm W}$, as $P_{\rm AP, max} = 30~{\rm W}$ results in infeasibility. It is observed that the total energy consumption decreases as $P_{\rm U, max}$ increases, as a higher transmit power budget enables UAVs to offload more computation tasks to the MEC server and significantly reduce the energy consumption of task computation. Furthermore, it is evident that the proposed design demonstrates superior performance over the two benchmarks. First, compared with the power allocation design, the proposed one leverages spatial degrees of freedom from multiple antennas via  directional beamforming, which simultaneously enhance both task offloading efficiency and target sensing performance. Second, compared with the straight flight design, the proposed one optimizes the UAV trajectories to reduce the propulsion energy consumption and improve the offloading link quality by reducing path loss. 

    \begin{figure}[!t]
		\centering
		\includegraphics[width=0.48\textwidth]{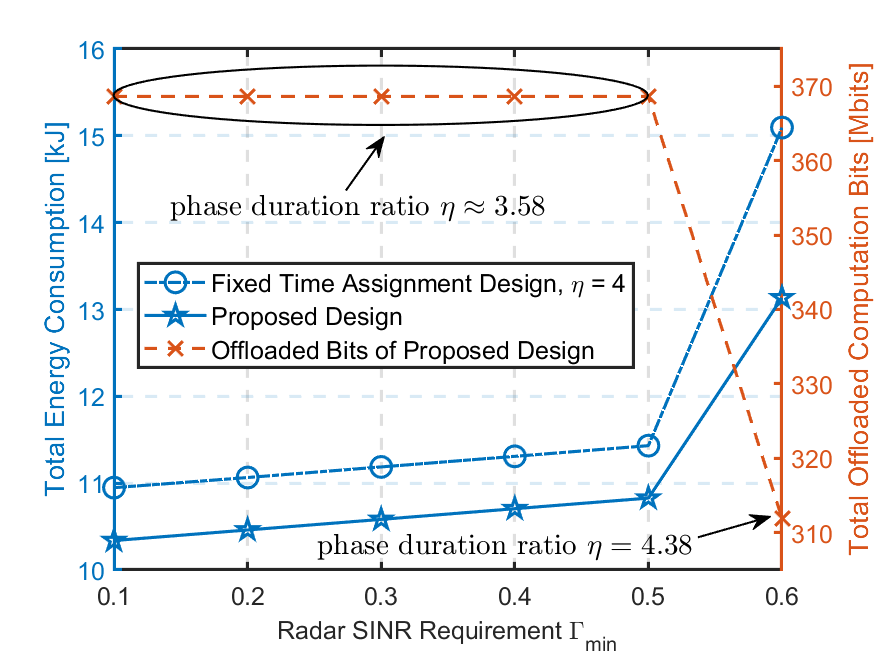}
		\caption{Comparison between the proposed and fixed time assignment designs.}
		\label{fig: gamma}
    \end{figure}
    

    Fig. \ref{fig: gamma} shows the total energy consumption obtained by the proposed design versus the radar SINR requirement $\Gamma_{\min}$. Here, we define the phase duration ratio between Phase 1 and Phase 2 as $\eta \triangleq (\sum_n t_0[n]) /  (\sum_n t_1[n])$. For comparison, the fixed time assignment design is  simulated with $\eta=4$. As observed, the proposed design obtains better system performance than the benchmark, which validates the effectiveness of time assignment. Increasing the duration of Phase 1 can enhance offloading efficiency (i.e., offloading more task bits or reducing transmit power consumption), but at the cost of shortening the duration of Phase 2 for the MEC server to execute these offloaded tasks, which may cause the server to operate in the high energy consumption range. The proposed design is able to dynamically adjust the time assignment  based on channel conditions, which well balance the trade-off between two phases. Besides, it is observed that the total energy consumption increases quasi-linearly with $\Gamma_{\min}\in [0.1, 0.5]$ but spikes dramatically with $\Gamma_{\min} \in [0.5, 0.6]$. To explore the reason, the corresponding offloaded task bits are provided and marked as the orange curve. Notably, the offloaded data and $\eta$ remain nearly unchanged with $\Gamma_{\min} \in [0.1, 0.5]$. This phenomenon indicates that for $\Gamma_{\min} \leq 0.5$, the system increases the transmit power of APs in Phase 2 to meet rising radar SINR requirements. However, for $\Gamma_{\min} > 0.5$, the stricter radar SINR requirement necessitates more sensing power and dedicated sensing time. Consequently, the duration of Phase 1 is shrunk accordingly (marked by increased $\eta$), which leads to more task bits processed locally with lower efficiency (marked by decreased offloaded bits).  Moreover, the gap in  total energy consumption between the two designs surges when $\Gamma_{\min}$ increases from $0.5$ to $0.6$, as the proposed design introduces one more degree of freedom (i.e., time assignment) to deal with the increment of $\Gamma_{\min}$.
    
    


     \begin{figure}[!t]
		\centering
        \includegraphics[width=0.48\textwidth]{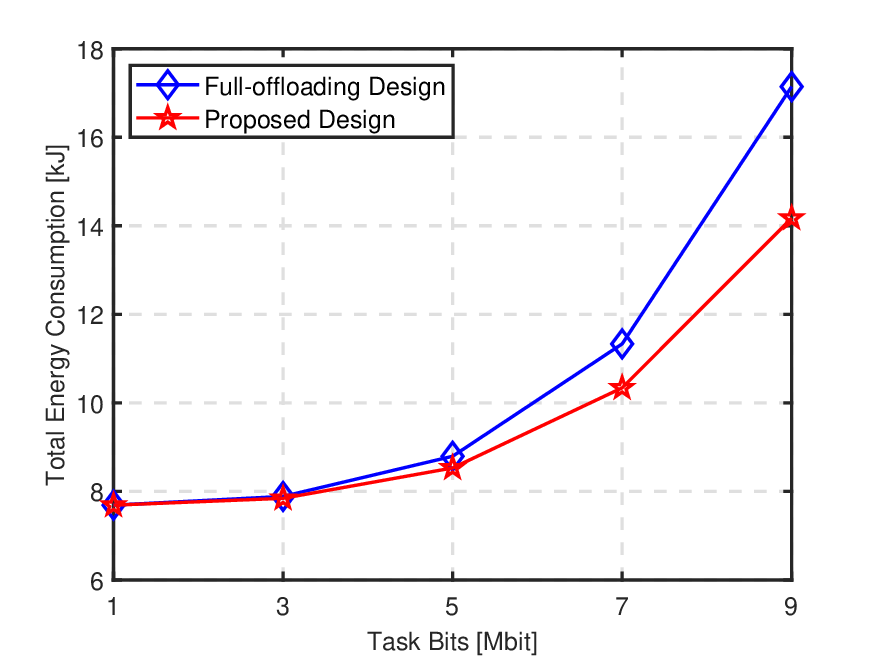}
		\caption{Comparison between the proposed and full-offloading designs.}
		\label{fig: Ik}
     \end{figure}

     Fig. \ref{fig: Ik} shows the total energy consumption versus the task bits $I_k[n]$. For comparison, we adopt the full-offloading design as a benchmark, where all computation tasks are offloaded to and executed by the MEC server, i.e., $l_{{\rm l}, k}[n] = 0$ $\forall k\in \mathcal{K}$ and $\forall n \in \mathcal{N}$. As observed, the total energy consumption of both designs increases with $I_k[n]$. This is because processing more task bits requires a higher CPU frequency,  leading to greater computational energy consumption. Furthermore, one can find that the proposed design outperforms the full-offloading design, and the performance gap grows larger  with $I_k[n]$. The reason is that by optimizing the offloading ratio, the proposed design can achieve balanced computation between UAVs and the MEC server. Especially for large tasks, the full-offloading design consumes excessive time for data transmission, leaving insufficient time for MEC computation. In contrast, the proposed design reduces the offloading overhead by local computation, thereby allocating more time for MEC computation and reducing  total energy consumption.

    \begin{figure}[!t]
		\centering
		\includegraphics[width=0.48\textwidth]{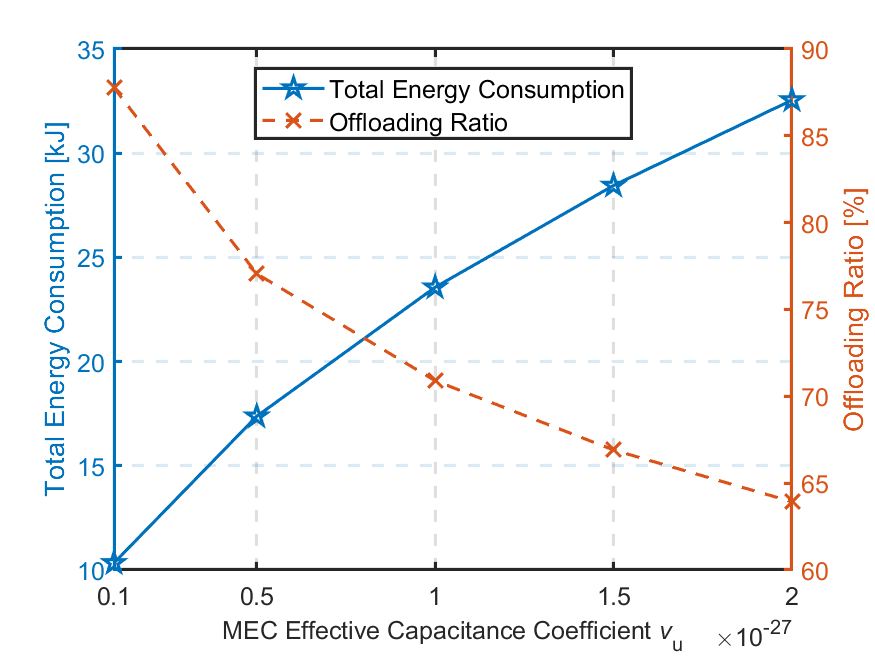}
		\caption{The total energy consumption and offloading ratio versus the MEC effective capacitance coefficient $v_{\rm u}$.}
		\label{fig: vu}
    \end{figure}

    Fig. \ref{fig: vu} presents the total energy consumption and offloading ratio (calculated by $(\sum_{k, n} l_{{\rm u}, k}[n]) /(\sum_{k, n} I_{k}[n])$) versus the MEC effective capacitance coefficient $v_{\rm u}$, with the UAV effective capacitance coefficient $v_{\rm l}$ being $10^{-26}$. One can find that the total energy consumption increases with  $v_{\rm u}$, while the offloading ratio decreases with $v_{\rm u}$. This occurs because a higher $v_{\rm u}$ escalates the computation cost of MEC server, which discourages task offloading. Notably, when $v_{\rm u}$ increases from $10^{-28}$ to $10^{-27}$, the offloading ratio drops sharply from $87.8\%$ to $70.9\%$, which suggests that $v_{\rm u}$ should be at least an order of magnitude lower than $v_{\rm l}$ to maintain the MEC deployment to be cost-effectively.

	\section{Conclusion}
    This paper has investigated the covert communications in a MEC-based networked ISAC system towards LAE. We considered a system comprising multiple APs coordinated by a MEC server and multiple UAVs in the presence of a target and several wardens. The UAVs offloaded tasks to the MEC via APs, while APs generated dual-functional waveforms to locate the target and interfere with wardens. We derived the closed-form expressions for the DEP at wardens under optimal decision threshold conditions. The total energy consumption of the considered system was minimized through a proposed AO-based algorithm subject to the QoS requirements of MEC service, the radar SINR requirements, the DEP requirements, and the causality of UAV trajectories. Extensive simulations validated the proposed algorithm and reveled the trade-offs among communication, sensing, and computation in LAE systems. 
    
    Our  future studies will explore more general application scenarios for LAE systems, including integrating novel B5G technologies into LAE and analyzing the impacts of imperfect channel state information on the system design.

\appendix
\subsection{Detailed Derivation Procedure for \eqref{psi_tilde} and \eqref{omega_tilde}}  

For convenience, we denote ${\bf W}_k[n] \triangleq {\bf w}_k[n] {\bf w}_k^H[n]$ and the derivative of $\Psi_k ({\bf u}_k[n])$ w.r.t. ${\bf u}_k[n]$ is rewritten as
\begin{flalign}
    \frac{{\rm d} \Psi_k ({\bf u}_k[n])}{{\rm d} {\bf u}_k[n]} &= \frac{{\rm d} {\rm Tr}\left( {\bf H}_k[n] {\bf W}_k[n] {\bf H}_k^H[n]   \right)}{{\rm d} {\bf u}_k[n] }  \\
    &= 2 {\rm Re} \left\{ \begin{bmatrix}
        {\rm Tr} \left( {\bf W}_k[n] {\bf H}_k^H[n] \frac{\partial {\bf H}_k[n] }{\partial x_k^{\rm U}[n] }   \right)    \\
        {\rm Tr} \left( {\bf W}_k[n] {\bf H}^H_k[n] \frac{\partial {\bf H}_k[n]}{\partial y_k^{\rm U}[n] }   \right)
    \end{bmatrix}     \right\}. \nonumber
\end{flalign}
where 
\begin{flalign*}
    &\frac{\partial {\bf H}_k[n]} {\partial x^{\rm U}_k[n]} = \left[ \frac{\partial {\bf H}_{1, k}[n]} {\partial x^{\rm U}_k[n]}; \frac{\partial {\bf H}_{2, k}[n]} {\partial x^{\rm U}_k[n]}; \ldots; \frac{\partial {\bf H}_{M, k}[n]} { \partial x^{\rm U}_k[n]}  \right] \\
    {\rm and}~&\frac{\partial {\bf H}_k[n]} {\partial y^{\rm U}_k[n]} = \left[ \frac{\partial {\bf H}_{1, k}[n]} {\partial y^{\rm U}_k[n]}; \frac{\partial {\bf H}_{2, k}[n]} {\partial y^{\rm U}_k[n]}; \ldots; \frac{\partial {\bf H}_{M, k}[n]} { \partial y^{\rm U}_k[n]}  \right] .
\end{flalign*}

First, we calculate the partial derivative of ${\bf H}_{m, k}[n]$ w.r.t. $x^{\rm U}_k[n]$ as follows:
\begin{flalign}
    &\frac{\partial {\bf H}_{m, k}[n] }{\partial x_k^{\rm U}[n]} = \sqrt{C_0} \frac{\partial \frac{ {\bf a}_{\rm R} \left( \theta_{m, k}[n]  \right) {\bf a}_{\rm U}^T\left( \theta_{m, k}[n]   \right)} {|| {\bf q}_k^{\rm U}[n]  - {\bf q}_m^{\rm AP} ||}  }{\partial x_k^{\rm U}[n]}  = \sqrt{C_0} \Bigg( \nonumber \\
    & \frac{ \left(x_m^{\rm AP} -x_k^{\rm U}[n]\right) {\bf a}_{\rm R} \left( \theta_{m, k}[n]  \right) {\bf a}_{\rm U}^T\left( \theta_{m, k}[n]   \right) }{ ||{\bf q}_k^{\rm U}[n] - {\bf q}_m^{\rm AP}  ||^3  }  + \frac{\partial {\bf a}_{\rm R} \left( \theta_{m, k}[n]  \right) }{\partial x_k^{\rm U}[n]} \times
    \nonumber \\
    &   \frac{{\bf a}_{\rm U}^T\left( \theta_{m, k}[n]   \right)}{|| {\bf q}_k^{\rm U}[n]  - {\bf q}_m^{\rm AP} ||} + \frac{{\bf a}_{\rm R} \left( \theta_{m, k}[n]  \right)}{|| {\bf q}_k^{\rm U}[n]  - {\bf q}_m^{\rm AP} ||}  \frac{\partial {\bf a}_{\rm U}^T\left( \theta_{m, k}[n]   \right) }{\partial x_k^{\rm U}[n]} \Bigg), 
    \label{hxk}
\end{flalign}
where $\partial {\bf a}_{\rm R}(\theta_{m, k}[n]) /\partial x_k^{\rm U}[n] $ and $\partial {\bf a}_{\rm U}(\theta_{m, k}[n]) /\partial x_k^{\rm U}[n] $ are given in \eqref{arx} and \eqref{aux}, respectively.
\begin{figure*}
\begin{flalign}
    & \frac{\partial {\bf a}_{\rm R} \left( \theta_{m, k}[n]  \right) }{\partial x_k^{\rm U}[n]} = \left[0, \ldots, j 2 \pi \frac{d}{\lambda}  (N_{\rm R}-1) \frac{H_k\left(x_m^{\rm AP} - x_k^{\rm U}[n]  \right)  }{|| {\bf q}_k^{\rm U}[n] - {\bf q}_m^{\rm AP}    ||^3}  e^{j 2 \pi \frac{d}{\lambda} \cos(\theta_{m,k}[n])(N_{\rm R}-1) }    \right]^T,  
    \label{arx}
    \\
    &\frac{\partial {\bf a}_{\rm U} \left( \theta_{m, k}[n]  \right) }{\partial x_k^{\rm U}[n]} = \left[0, \ldots, j 2 \pi \frac{d}{\lambda}  (N_{\rm U}-1) \frac{H_k\left(x_m^{\rm AP} - x_k^{\rm U}[n]  \right)  }{|| {\bf q}_k^{\rm U}[n] - {\bf q}_m^{\rm AP}    ||^3}  e^{j 2 \pi \frac{d}{\lambda} \cos(\theta_{m,k}[n])(N_{\rm U}-1) }    \right]^T.  
    \label{aux}
\end{flalign}
\hrule
\end{figure*}

Next, we calculate the partial derivative of ${\bf H}_{m, k}[n]$ w.r.t. $y_k^{\rm U}[n]$. Notably, due to the inherent symmetry between $(x_k^{\rm U}[n] -x_m^{\rm AP})$ and $(y_k^{\rm U}[n] - y_m^{\rm AP} )$ in ${\bf H}_{m,k}[n]$, the expression for $\partial {\bf H}_{m, k}[n] / \partial y_k^{\rm U}[n]$ can be derived analogously by substituting $x_k^{\rm U}[n]$ and $x_m^{\rm AP}$ in equation \eqref{hxk} with $y_k^{\rm U}[n]$ and $y_m^{\rm AP}$, receptively.

Following the above calculation, we derive the expression for ${\rm d} \Psi_k ({\bf u}_k[n]) / {\rm d} {\bf u}_k[n]$ in equation \eqref{psi_tilde}. The derivation process of the expression for ${\rm d} \Omega_{l, k} ({\bf u}_k[n]) / {\rm d} {\bf u}_k[n]$ is analogous to that of ${\rm d} \Psi_k ({\bf u}_k[n]) / {\rm d} {\bf u}_k[n]$. Therefore, we omit the detailed steps and directly present its expression as follows:
\begin{flalign}
     \frac{{\rm d} \Omega_{l, k}({\bf u}_k[n])  }{{\rm d} {\bf u}_k[n]} = 2 {\rm Re} \left(
     \begin{bmatrix}
     {\bf h}_{l, k}^H[n] {\bf W}_k[n] \frac{\partial {\bf h}_{l, k}[n] }{\partial x_k^{\rm U}[n] } 
     \\
     {\bf h}_{l, k}^H[n] {\bf W}_k[n] \frac{\partial {\bf h}_{l, k}[n] }{\partial y_k^{\rm U}[n] } 
     \end{bmatrix}
     \right),
\end{flalign}
where 
\begin{flalign*}
    \frac{\partial {\bf h}_{l, k}[n] }{\partial x_k^{\rm U}[n] } \triangleq&~ \sqrt{C_0} \Bigg( \frac{x_l^{\rm W} - x_k^{\rm U}[n]   }{||{\bf q}_k^{\rm U}[n] - {\bf q}_l^{\rm W}  ||^3 } {\bf a}_{\rm U} \left( \phi_{l,k}[n]  \right)
    + \\
    &
     \frac{\partial  {\bf a}_{\rm U} \left( \phi_{l,k}[n]  \right)}{ || {\bf q}_k^{\rm U}[n] - {\bf q}_l^{\rm W} || \partial x_k^{\rm U}[n] }
    \Bigg) \\
    {\rm and }~
    \frac{\partial {\bf h}_{l, k}[n] }{\partial y_k^{\rm U}[n] } \triangleq &~ \sqrt{C_0} \Bigg( \frac{y_l^{\rm W} - y_k^{\rm U}[n]   }{||{\bf q}_k^{\rm U}[n] - {\bf q}_l^{\rm W}  ||^3 } {\bf a}_{\rm U} \left( \phi_{l,k}[n]  \right)
    + \\
    &
     \frac{\partial  {\bf a}_{\rm U} \left( \phi_{l,k}[n]  \right)}{ || {\bf q}_k^{\rm U}[n] - {\bf q}_l^{\rm W} || \partial y_k^{\rm U}[n] }
    \Bigg)
\end{flalign*}
with $\partial {\bf a}_{\rm U}(\phi_{l, k}[n]) / \partial x_k^{\rm U}[n] $ and $\partial {\bf a}_{\rm U}(\phi_{l, k}[n]) / \partial y_k^{\rm U}[n] $ giving in \eqref{aulkx} and \eqref{aulky}, receptively.
\begin{figure*}
\begin{flalign}
    & \frac{\partial {\bf a}_{\rm U}(\phi_{l, k}[n])}{\partial x_k^{\rm U}[n]} = \left[0, \ldots, j 2 \pi \frac{d}{\lambda}  (N_{\rm U}-1) \frac{\left( H_l^{\rm W} - H_k \right)\left(x_l^{\rm W} - x_k^{\rm U}[n]  \right)  }{|| {\bf q}_k^{\rm U}[n] - {\bf q}_l^{\rm W}    ||^3}  e^{j 2 \pi \frac{d}{\lambda} \cos(\phi_{l, k}[n]))(N_{\rm U}-1) }    \right]^T,  
    \label{aulkx}
    \\
    &\frac{\partial {\bf a}_{\rm U}(\phi_{l, k}[n])}{\partial y_k^{\rm U}[n]} = \left[0, \ldots, j 2 \pi \frac{d}{\lambda}  (N_{\rm U}-1) \frac{\left(  H_l^{\rm W} - H_k \right)\left(y_l^{\rm W} - y_k^{\rm U}[n]  \right)  }{|| {\bf q}_k^{\rm U}[n] - {\bf q}_l^{\rm W}    ||^3}  e^{j 2 \pi \frac{d}{\lambda} \cos(\phi_{l, k}[n])(N_{\rm U}-1) }    \right]^T.  
    \label{aulky}
\end{flalign}
\hrule
\end{figure*}

\end{document}